\documentclass[aps,prd,superscriptaddress,nofootinbib,amsmath,amsfonts,preprintnumbers,groupedaddress,showpacs,10pt,english]{revtex4-1}
\usepackage{amsmath}
\usepackage{amssymb}
\usepackage{babel}
\usepackage{wrapfig}
\usepackage{cancel}
\newcommand{\nn}{\nonumber \\}
\newcommand{\e}{\mathrm{e}}

\makeatletter

\usepackage{array,multirow,graphicx}
\usepackage{dcolumn}
\usepackage{newlfont}
\usepackage{bm}
\usepackage[colorlinks,citecolor=blue,urlcolor=blue,linkcolor=blue]{hyperref}
\usepackage[figtopcap]{subfigure}
\usepackage{color}

\begin{document}

\date{\today}

\title{Analytic charged BHs in $f(R)$ gravity}

\author{G.~G.~L.~Nashed}
\email{nashed@bue.edu.eg}
\affiliation {Centre for Theoretical Physics, The British University, P.O. Box
43, El Sherouk City, Cairo 11837, Egypt}

%\author{S.D. Odintsov}
%\email{odintsov@ieec.uab.es}
%\affiliation{Institut de Ci\`encies de l'Espai (ICE-CSIC/IEEC),
%Campus UAB, c. Can Magrans s/n, 08193, Barcelona, Spain}
%\affiliation{Instituci\'o Catalana de Recerca i Estudis Avan\c{c}ats (ICREA),
%Barcelona, Spain}
%\affiliation{Tomsk State Pedagogical University, 634061 Tomsk, Russia}

\author{S.~Nojiri}
\email{nojiri@gravity.phys.nagoya-u.ac.jp}
\affiliation{Kobayashi Maskawa Institute for the Origin of Particles and the Universe, Nagoya University, Nagoya 464-8602, Japan}

\begin{abstract}
In this article, we seek exact charged spherically symmetric black holes (BHs) with considering $f(R)$ gravitational theory.
This BH is characterized by convolution and error functions.
Those two functions depend on a constant of integration which is responsible to make such a solution deviate from Einstein's general relativity (GR).
The error function which constitutes the charge potential of the Maxwell field depends on the constant of integration and when this constant is vanishing
we cannot reproduce the Reissner-Nordstr\"om BH in the lower order of $f(R)$.
This means that we cannot reproduce Reissner-Nordstr\"om BH in lower-order-curvature theory, i.e., in GR limit $f(R)=R$, we cannot get the well known charged BH.
We study the physical properties of these BHs and show that it is asymptotically approached as a flat spacetime or approach AdS/dS spacetime.
Also, we calculate the invariants of the BHS and show that the singularities are milder than those of BHs in GR.
Additionally, we derive the stability condition through the use of geodesic deviation.
Moreover, we study the thermodynamics of our BH and show the effect of the higher-order-curvature theory.
Finally, we study the stability analysis by using the odd-type mode and show that all the BHs are stable and have radial speed equal to one.
\end{abstract}


%\begin{center}
\maketitle
\section{Introduction}
%\end{center}
Despite the simplicity, elegantly and powerfully of GR, which is considered as a classical theory, and despite its successes in the prediction of the shadow of BH,
which confirmed by horizon telescope \cite{Akiyama:2019cqa,Akiyama:2019eap,Ayzenberg:2018jip}, and its success with observation, like the bending of light,
perihelion precession of mercury, the gravitational redshift of radiation from distant stars \cite{Will:2005yc,Will:2005va,Yunes:2013dva}, it has unsolved problems.
Among these issues, is the problem of dark energy and dark matter which confirmed by recent observation
\cite{Milgrom:1983pn,Bekenstein:1984tv,Milgrom:2003ui,Perlmutter:1998np,Riess:1998cb}, the problem of singularities
\cite{PhysRevLett.14.57,PhysRevD.14.2460,Christodoulou:1991yfa}, the problem of quantum gravity \cite{PhysRevLett.77.3288,Ashtekar:2006rx}.
All these issues make physicists demand another theory that is eligible and able to overcome such problems and permit GR in the lower energy.
Therefore, a satiate modified gravitational theory has been constructed that can overcome the issues that GR cannot solve.
Later, physicists put constraints that any modified gravity must bypass like: its consistency with the solar system, free from ghost modes,
able to describe the observations in which GR could not solve, and must not create a fifth force in local physics.
There are many modified theories that satisfy such benchmark and are categorized as:
\begin{itemize}
\item[a)] Theories in which their Lagrangian involve higher-order curvature like $f(R)$, \cite{Nojiri:2010wj,Capozziello:2006dj},
Lanczos-Lovelock models \cite{PhysRev.39.716,Lanczos:1938sf,Lovelock:1971yv,Padmanabhan:2013xyr,Dadhich:2015ivt} etc.
\item[b)] Higher-dimensional theories which change the equation of motions of 4-dimension because of the higher-order
correction of the energy-momentum tensor \cite{Shiromizu:1999wj,Awad2017,Harko:2004ui,Carames:2012gr,Haghani:2012zq,Chakraborty:2014xla}.
\item[c)] Scalar theories of gravity like Brans-Dicke theory and the more general Horndeski models \cite{PhysRev.124.925,Horndeski:1974wa,Sotiriou:2013qea,Babichev:2016rlq}.
\end{itemize}

In this study, we take $f(R)$ gravitational theory which is considered as the excellent modification of the Einstein GR at the most recent time.
Among many things that make $f(R)$ gravity be the best modification of GR is its simplified Lagrangian which contains an arbitrary function whose lowest order coincides with Einstein's theory.
The $f(R)$ gravity is an elegant theory to describe an expanded accelerated universe which is supported from recent observations \cite{Riess:2004nr,Riess:1998cb}.
Moreover, it is supposed that a special form of $f(R)$ can mimic some curvature modifications that come from the quantum theory of gravity.
The justifications of invented modified theories of GR are to construct complete systems of gravitational models to cure the issues appearing in GR.
These systems are considered in various topics of gravity like astrophysics, cosmology and the output of these studies construct as a suitable extension
of the Einstein GR \cite{Cognola:2005de,PhysRevD.82.064033,PhysRevD.88.124036}.
Recently, the $f(R)$ gravity theory gains much attentions which generalize the gravitational Lagrangian of GR \cite{RevModPhys.82.451,Hendi:2009sw,Capozziello:2011et,PhysRevD.39.3159}.
There are two methods to obtain the equation of motions of $f(R)$ gravitational theory, the first through the use of metric structure and the other through the use of
the Palatini method in which the connection and metric are taken as independent variables \cite{Antoniadis:2020dfq}.
There are many viable applications of $f(R)$ in the domain of cosmology as well as astrophysics \cite{PhysRevD.75.083504,PhysRevD.79.103521,PhysRevD.76.104043,Starobinsky:2007hu}
and in a specific case, $f(R)$ can reduce to GR case \cite{Capozziello:2015hra}.

There are many spherically symmetric BH solutions derived in $f(R)$ \cite{PhysRevD.74.064022,2018EPJP..133...18N,2018IJMPD..2750074N,Nashed2002521,Nashed:2018piz}.
Moreover, novel spherically symmetric BH solutions have been obtained in the context of $f(R)$ \cite{Elizalde:2020icc,Nashed:2019yto,Nashed:2019tuk}.
It is well known that higher-order curvature terms play an ingredient role in strong gravitational background in local objects of $f(R)$.
In this context, many physicists concentrate on the study of spherically symmetric
BHs \cite{Sultana:2018fkw,Canate:2017bao,Nashed:2009hn,Yu:2017uyd,Canate:2015dda,Kehagias:2015ata,PhysRevD.82.104026,delaCruzDombriz:2009et}.
This exposition aims to derive original spherically symmetric charged BH solutions in $f(R)$ gravity devoid of any constraints on the form of $f(R)$ nor the Ricci scalar
and study their contingent physics.
{ In the frame of $f(R)$-gravity, a Lagrangian approach has been developed to study dynamics of spherically symmetric metrics.
The Euler-Lagrange equations are obtained and solved in the case of constant curvature
$R = R_0$ recovering the standard Schwarzschild-de Sitter solution of GR \cite{Capozziello:2012iea}.}

The paper is structured as: In Sec.~\ref{S2}, we give a brief construction of $f(R)$ gravity.
In Sec.~\ref{S3}, we employ the charged equation of motions of $f(R)$ to a spherically symmetric spacetime having two unknown metric potentials
and derive a system of non-linear differential equations.
These systems have four unknown functions, two of the metric potentials, one from the gauge potential of the charge and the derivative of $f(R)$.
To be able to solve this system and to put it in a closed system, we assume a form of the derivative of $f(R)$, which depends on a constant of integration and solve the system.
The BH solution is characterized by convolution and error functions and they become constant functions when the constant of integration
that appears in the derivative function of $f(R)$ is equal to zero and in that case, we have a GR BHs.
Therefore, the convolution and error functions appear like the effect of higher-order curvature that characterizes $f(R)$ gravity.
Moreover, we study the physics of these BHs by giving the asymptote form of the convolution and error functions up to certain order and showing the effect of the higher-order curvature.
We examine the singularities of the BHs by calculating the invariants of the Kretschmann scalar, the Ricci tensor square, and the Ricci scalar and investigate
the effect of higher-order curvature on these invariants by showing that the singularity of our BHs is milder than the GR BHs.
Also in Sec.~\ref{S3}, we calculate the geodesic deviation and give the condition of stability.
In Sec.~\ref{S4}, we show the validity of the first law of thermodynamics and also calculate some thermodynamic quantities like the Hawking temperature, entropy,
the Gibbs free energy, and quasi-local energy of these BHs.
We reserve the final section for the discussion and conclusion of this study.

\section{Fundamentals of $f(R)$ gravitational theory}\label{S2}
We are going to study a 4-dimensional Lagrangian of $f(R)$ with $f$ being an arbitrary function.
The Lagrangian of $f(R)$ is given by (cf. \cite{Carroll:2003wy,1970MNRAS.150....1B,Nojiri:2003ft,Capozziello:2003gx,Capozziello:2011et,Nojiri:2010wj,Nojiri:2017ncd,Capozziello:2002rd}):
\begin{eqnarray}
\label{a2}
{\mathop{\mathcal{ L}}}:=\frac{1}{2\kappa} \int d^4x \sqrt{-g}~f(R)+\int d^{4}x~\sqrt{-g}~\mathcal{L}_\mathrm{em}\,.
\end{eqnarray}
Here $\kappa$ and $g$ are the Newtonian gravitational constant and the determinant of the metric.
The Lagrangian of electromagnetic field $\mathcal{L}_\mathrm{em}$ is defined as $\mathcal{L}_\mathrm{em}= {F}$ where ${F} = d{\xi}$ and ${\xi}={\xi}_{\beta}dx^\beta$ is the
electromagnetic Maxwell gauge potential 1-form \cite{Capozziello:2012zj}.

Making the variations principle w.r.t. the metric tensor $g_{\alpha \beta}$ to the Lagrangian (\ref{a2}) we get the equation of motions of
$f(R)$ gravitational theory in the form \cite{2005JCAP...02..010C}
%,Koivisto:2005yc }
%\newpage
\begin{eqnarray}
\label{f1}
{\mathop{\mathcal{ I}}}_{\mu \nu}={ R}_{\mu \nu} f_R-\frac{1}{2}g_{\mu \nu}f(R)+ \left[ g_{\mu \nu}\Box -\nabla_\mu \nabla_\nu \right]f_R
+\frac{1}{2}\kappa {T_\mathrm{(em)}}_{\mu\nu} \equiv 0 \,,
\end{eqnarray}
with $\Box$ being the d'Alembertian operator, $\displaystyle f_R=\frac{d{f}}{{d{R}}}$, and ${T_\mathrm{(em)}}^\nu_\mu$ is the
energy-momentum tensor of the Maxwell field defined as
\begin{equation}
\label{en11}
{T_\mathrm{(em)}}^{\ \nu}_\mu={F}_{\mu \alpha}{F}^{\nu\alpha}-\frac{1}{4} \delta_\mu^{\ \nu} {F} \,,
\end{equation}
where ${F}={F}_{\mu \nu}{F}^{\mu \nu}$.
Moreover, the variation of Eq.~(\ref{a2}) w.r.t. the gauge potential 1-form, ${\xi}_{\mu}$, gives:
\begin{eqnarray}
\label{q8b}
&&\partial_\alpha\left( \sqrt{-g} {F}^{\mu \nu} \right)=0\, .
\end{eqnarray}

The trace of the field equations ~(\ref{f1}), takes the form:
\begin{eqnarray}
\label{f3}
{\mathop{\mathcal{E}}}=3\Box f_R+Rf_R-2f(R)\equiv0 \,.
\end{eqnarray}
From Eq.~(\ref{f3}) one can obtain $f(R)$ in the form:
\begin{eqnarray}
\label{f3s}
f(R)=\frac{1}{2} \left[ 3\Box f_R+Rf_R \right]\,.
\end{eqnarray}
Using Eq.~(\ref{f3s}) in Eq.~(\ref{f1}) we get \cite{Kalita:2019xjq}
\begin{eqnarray}
\label{f3ss}
I_{\mu \nu}=R_{\mu \nu} f_R-\frac{1}{4}g_{\mu \nu}Rf_R+\frac{1}{4}g_{\mu \nu}\Box f_R -\nabla_\mu \nabla_\nu f_R
+\frac{1}{2}\kappa {T_\mathrm{(em)}}_{\mu\nu}=0 \,.
\end{eqnarray}
We will apply the field equations (\ref{f3}), (\ref{f3ss}), and (\ref{en11}) to a spherically symmetric spacetime with two unequal unknown functions in the next section.
%%%%%%%%%%%%%%%%%%%%%%%%%%%%%%%%%%% Section 3 %%%%%%%%%%%%%%%%%%%%%%%%%%%%%%%%%%%%%%%%
\section{Spherically symmetric black hole solutions}\label{S3}
%%%%%%%%%%%%%%%%%%%%%%%%%%%%%%%%%%%%%%%%%%%%%%%%%%%%%%%%%%%%%%%%%%%%%%%%%%%%%%%%%%%%%%
Let us take the following line-element
%%%%%%%%%%%%%%%%%%%%%%%%%%%%%%%%%%% Section 3 %%%%%%%%%%%%%%%%%%%%%%%%%%%%%%%%%%%%%%%%
%\subsection{Spherically symmetric solution}
%Assuming the spherically-symmetric the line-element to be in the form:
\begin{eqnarray}
\label{met12}
& & ds^2=-\mu(r)dt^2+\frac{dr^2}{\nu(r)}+r^2 \left( d\theta^2+\sin^2d\phi^2 \right) \,,
\end{eqnarray}
which describes a spherically symmetric a spacetime with $\mu(r)$ and $\nu(r)$ which are two functions of the radial coordinate $r$ to be determined from the equation of motions.
 From Eq.~(\ref{met12}), the Ricci scalar takes the form:
\begin{eqnarray}
\label{Ricci}
{R}(r)=\frac{r^2\nu\mu'^2-r^2\mu\mu'\nu'-2r^2\mu \nu\mu''-4r\mu[\nu\mu'-\mu \nu']+4\mu^2(1-\nu)}{2r^2\mu^2}\,,
\end{eqnarray}
where $\mu\equiv \mu(r)$, $\nu\equiv \nu(r)$, $\mu'=\frac{d\mu}{dr}$, $\mu''=\frac{d2\mu}{dr^2}$, and $\nu'=\frac{d\nu}{dr}$.
Applying the equation of motions (\ref{q8b}), (\ref{f3}) and (\ref{f3ss}) to the line-element (\ref{met12}) using Eq.~(\ref{Ricci}), we get the following non-linear differential equations:
\begin{align}
{\mathop{\mathcal{ I}}}_t{}^t=&\frac{1}{8r^2\alpha^2}\left\{r^2 \left[ \nu {F}_1\alpha'^2-3\alpha {F}_1\nu'\alpha'-2\alpha\nu {F}_1\alpha''-2\alpha^2{F}_1\nu''
 -3\alpha\nu \alpha'{F}'_1-2\alpha^2\nu'{F}'_1+2\nu \alpha^2{F}'_1 \right] \right. \nonumber\\
& \left. -4r\alpha\nu \left[ {F}_1\alpha'-\alpha {F}'_1 \right] -4\alpha^2{F}_1 \left[ 1-\nu \right]'-8\alpha r^2 {\xi}'^2 \right\}=0\,,\nonumber\\
{\mathop{\mathcal{ I}}}_r{}^r=&\frac{1}{8r^2\alpha^2} \left\{r^2 \left[ \nu {F}_1\alpha'^2-3\alpha {F}_1\nu'\alpha'-2\alpha\nu {F}_1\alpha''-2\alpha^2{F}_1\nu''+\alpha\nu \alpha'{F}'_1-2\alpha^2\nu'{F}'_1-6\nu \alpha^2{F}''_1 \right] \right. \nonumber\\
& \left. +4r\alpha\nu \left[ {F}_1\alpha'+\alpha{F}'_1 \right]-4\alpha^2{F}_1 \left[ 1-\nu \right]-8\alpha r^2 {\xi}'^2 \right\}=0\,,\nonumber\\
{\mathop{\mathcal{ I}}}_\theta{}^\theta=& {\mathop{\mathcal{ I}}}_\phi{}^\phi=\frac{1}{8r^2\alpha^2}
\left\{r^2 \left[3\alpha {F}_1\nu'\alpha'+2\alpha\nu {F}_1\alpha''+2\alpha^2{F}_1\nu''-\nu {F}_1\alpha'^2+\alpha\nu \alpha'{F}'_1+2\alpha^2\nu'{F}'_1+2\nu \alpha^2{F}''_1 \right] \right. \nonumber\\
& \left. -4r\alpha^2\nu {F}'_1+4\alpha^2{F}_1 \left[ 1-\nu \right]+8\alpha r^2 {\xi}'^2 \right\}=0\,,\nonumber\\
{\mathop{\mathcal{ I}}}=&\frac{1}{2r^2\alpha^2} \left\{ r^2 \left[ 6\alpha^2\nu'{F}'_1-3\alpha {F}_1\nu'\alpha'-2\alpha\nu {F}_1\alpha''-2\alpha^2{F}_1\nu''
+\nu {F}_1\alpha'^2+3\alpha\nu \alpha'{F}'_1+6\nu \alpha^2{F}''_1 \right] \right. \nonumber\\
& \left. +4r\alpha \left[ 3\alpha\nu {F}'_1-{F}_1\nu \alpha'-2{F}_1\alpha\nu' \right] +4\alpha^2{F}_1 \left[ 1-\nu \right]-4r^2\alpha^2f(r) \right\}=0\,,
\label{feq}
\end{align}
where $\alpha(r)=\frac{\mu(r)}{\nu(r)}$ and ${F}_1\equiv{F}_1(r)=\frac{d{f}({R}(r))}{d{R}(r)}$, ${F}'_1=\frac{d{F}_1(r)}{dr}$, ${F}''_1=\frac{d^2{F}_1(r)}{dr^2}$, ${F}'''_1=\frac{d^3{F}_1(r)}{dr^3}$.
The non-vanishing components of the Maxwell field has the form
\begin{equation}
\label{feqc}
\frac{{\xi}' \left[ r\alpha'-4\alpha \right]-2r\alpha {\xi}''}{2r\alpha^2}=0\, .
\end{equation}

When we exclude the trace part, Eqs.~(\ref{feq}) can have the form:
\begin{align}
\label{E1n}
0=& r^2\left[\nu{F}_1{\alpha'}^2 - 3\alpha{F}_1\nu'\alpha' - 2\alpha\nu{F}_1\alpha'' - 2\alpha^2{F}_1\nu'' - 3\alpha\nu\alpha'{F}'_1 - 2\alpha^2\nu'{F}'_1
+ 2\nu\alpha^2{F}''_1 \right] \nn
& - 4r\alpha\nu \left[ {F}\alpha' - \alpha{F}' \right] - 4\alpha^2{F} \left[1 - \nu \right]-8\alpha r^2 {\xi}'^2 \, , \\
\label{E2n}
0=&r^2 \left[ \nu{F}_1{\alpha'}^2 - 3\alpha{F}_1\nu'\alpha' - 2\alpha\nu{F}_1\alpha'' - 2\alpha^2{F}_1\nu'' + \alpha\nu\alpha'{F}'_1 - 2\alpha^2\nu'{F}'_1
 - 6\nu\alpha^2{F}''_1 \right] \nn
& +4r\alpha\nu \left[{F}_1\alpha'+ \alpha{F}'_1 \right] - 4\alpha^2{F}_1 \left[1 - \nu\right]-8\alpha r^2 {\xi}'^2 \, , \\
\label{E3n}
0=& r^2 \left[ - \nu{F}_1{\alpha'}^2 + 3\alpha{F}_1\nu'\alpha' + 2\alpha\nu{F}_!\alpha'' + 2\alpha^2{F}_1\nu'' + \alpha\nu\alpha'{F}'_1 + 2\alpha^2\nu'{F}'_1
+ 2\nu\alpha^2{F}''_1 \right] \nn
& - 4r\alpha^2\nu{F}'_1 + 4\alpha^2{F}_1 \left[1 - \nu\right]+8\alpha r^2 {\xi}'^2 \, .
\end{align}
Using Eqs.~(\ref{E1n}) and (\ref{E2n}), i.e., [(\ref{E1n}) minus (\ref{E2n})], we get
\begin{equation}
\label{E4n}
0 = r^2 \left[ - 4\alpha\nu\alpha'{F}' + 8\nu\alpha^2 {F}'' \right] - 8r \alpha\alpha'\nu {F} \, .
\end{equation}
Moreover, Eqs.~(\ref{E1n}) and (\ref{E3n}), i.e., [(\ref{E1n}) plus (\ref{E3n})] give:
\begin{equation}
\label{E6n}
0 = - 2 r^2 \alpha\nu\alpha'{F}' + 4r^2\nu\alpha^2 {F}'' - 4r \alpha\alpha' \nu{F} \,.
\end{equation}
Careful check can show that Eqs.~(\ref{E6n}) and (\ref{E4n}) are coincide and therefore we get
two equations from Eqs.~(\ref{E1n}), (\ref{E2n}) and (\ref{E3n}) which are independent.
Using the above information, we can say that Eq.~(\ref{E1n}) is equal to minus Eq.~(\ref{E2n}) minus two time Eq.~(\ref{E3n}).
Thus, Eqs.~(\ref{E1n}) and Eq.~(\ref{E6n}) can be chosen as independent equations.
Due to the fact that we have four unknown functions $\nu$, $\alpha$, ${\xi}$, and ${F}$, we cannot determine one function.

Let us discuss some special cases of Eq.~(\ref {E4n}) or Eq.~(\ref{E6n}). we can get the Reissner-Nordstr\"om solution by supposing,
\begin{equation}
\label{ES1}
\alpha=1 \, .
\end{equation}
Assuming $\nu\neq 0$ we can get from Eq.~(\ref{E4n})
\begin{equation}
\label{ES2n}
{F}''_1=0 \, , \quad \mbox{which gives,} \quad {F}_1={F}_2 + {F}_3 r \, .
\end{equation}
Using Eq.~(\ref{ES2n}) in Eq.~(\ref{E1n}) we get
\begin{align}
\label{ES3n}
0=& r^2\left[ - 2 {F}_1\nu'' - 2 \nu'{F}'_1 \right] + 4r \nu {F}'_1 - 4{F}_1 \left[1 - \nu \right]-8 r^2 {\xi}'^2 \nn
=& - 2 r^2 \left( {F}_2 + {F}_3 r \right) \nu'' -2 r^2 {F}_3 \nu' + 4 \left( {F}_2 + 2 {F}_3 r \right) \nu
 - 4 \left( {F}_2 + {F}_3 r \right)-8r^2 {\xi}'^2 \, .
\end{align}
Assuming ${F}_3=0$, Eq.~(\ref{ES3n}) gives
\begin{equation}
\label{ES4n}
0= {F}_2 [r^2 \nu''- 2 \nu + 2]+4 r^2 {\xi}'^2 \, .
\end{equation}
From Eq.~(\ref{ES4n}) after using Eq.~(\ref{feqc}) we get the following solution
\begin{equation}
\label{ES5n}
\nu= 1 + \frac{\nu_0}{r} + \nu_1 r^2-\frac{\nu_2}{{F}_2 r^2}\,, \qquad {\xi}=\frac{\nu_2}{ r}\, ,
\end{equation}
where $\nu_0$ $\nu_1$ and $\nu_2$ are integration constants.
Equation~(\ref{ES5n}) is the well-known Reissner-Nordstr\"om-(anti-)de Sitter space-time.

Moreover, we study the case ${F}_2=0$ in which Eq.~(\ref{ES3n}) gives,
\begin{equation}
\label{ES6n}
0 =- 2 r^2 {F}_3 r \nu'' -2 r^2 {F}_3 \nu' + 8 {F}_3 r \nu
 - 4 {F}_3 r -8r^2 {\xi}'^2\, .
\end{equation}
The solution of Eq.~(\ref{ES6n}) together with Eq.~(\ref{feqc}) have the following solution
\begin{equation}
\label{ES7n}
\nu= \frac{1}{2} + \frac{{\tilde \nu}_0 r^2}{20} + \frac{{\tilde \nu}_1}{ r^{2}}+ \frac{{\tilde \nu}_2}{2 r^{3}} \,, \qquad \qquad {\xi}=\frac{\sqrt{5{F}_3 \tilde \nu_2}}{2 r}\, ,
\end{equation}
where ${\tilde \nu}_0$, ${\tilde \nu}_1$ and ${\tilde \nu}_2$ are constants.
Equation~(\ref{ES7n}) corresponds to the solution derived before in \cite{Nashed:2019tuk,Elizalde:2020icc}.

For the general case, i.e., when ${F}_2$ and ${F}_3$ are not vanishing and for small $r$ then ${F}_2$ term in Eq.~(\ref{ES3n})
will be dominates and the solution must has the form given in Eq.~(\ref{ES5n}).
When $r$ is large then ${F}_3$ term in Eq.~(\ref{ES3n}) will be dominated and the solution has the form given by Eq.~(\ref{ES7n}).
Thus, there should be a solution that connects the two solutions given by Eqs.~(\ref{ES5n}), for a small $r$ region, and
(\ref{ES7n}) for large $r$ region.

Because we are studying a spherical symmetry case, we assume $f(R)=f(r)$.
The above system of differential equations, Eqs.~(\ref{feq}) and (\ref{feqc}), are five non-linear differential equations in four unknowns, ${F}_1$, $\nu$, $\alpha$, and ${\xi}$.
Thus, we cannot solve such a system unless we adjust the unknown with the number of differential equations.
{ The only way to solve the above system is to exclude the trace equation, i.e., we must not take into account the differential equation ${\mathop{\mathcal{ I}}}$
from Eq.~(\ref{feq}) and in that case, the solution of the system becomes:}
\begin{align}
\label{ass1}
\nu(r)=&\frac{\e^{\frac{3a_1}{2r^2}}}{r} \left\{{H}a_2+{H}_1r^3a_3+2{H}_1r^3\int\frac{\e^{\frac{-3a_1}{2r^2}}{H} \left[ r^2+13a_1 \right]}
{r \left( r^2+a_1 \right) \left[ \left(2a_1{H}_2 - 3r^2{H} \right) {H}_1-2a_1{H}{H}_3 \right]}dr \right. \nonumber\\
& \left. -2{H}\int\frac{\e^{\frac{-3a_1}{2r^2}}r^2 {H}_1 \left[ r^2+13a_1 \right]}{ \left( r^2+a_1 \right)
\left[ \left( 2a_1{H}_2 - 3r^2{H} \right) {H}_1-2a_1{H}{H}_3 \right]}dr \right\}\,,\nonumber\\
\alpha(r) =& \e^{\frac{-3a_1}{r^2}}\,, \qquad \mu(r)=\alpha(r)\nu(r)\,, \qquad {\xi}=a_0+\sqrt{\pi}\,\mathrm{erf} \left[\frac{\sqrt{6a_1}}{2r} \right]\,, \qquad {F}_1=1+\frac{a_1}{r^2}\,,
\end{align}
where $a_0$, $a_1$, $a_2$, and $a_3$ are constants.
Equation~(\ref{ass1}) shows that when $a_1=0$, we return to GR case.
Here, ${H}=\mathrm{HeunC}\left(\frac{3}{2},\frac{3}{2},0,\frac{3}{8},\frac{9}{8},-\frac{a_1}{r^2}\right)$,
${H}_1=\mathrm{HeunC}\left(\frac{3}{2},-\frac{3}{2},0,\frac{3}{8},\frac{9}{8},-\frac{a_1}{r^2}\right)$,
${H}_2=\mathrm{HeunCPrime}\left(\frac{3}{2},\frac{3}{2},0,\frac{3}{8},\frac{9}{8},-\frac{a_1}{r^2}\right)$,
${H}_3=\mathrm{HeunCPrime}\left(\frac{3}{2},-\frac{3}{2},0,\frac{3}{8},\frac{9}{8},-\frac{a_1}{r^2}\right)$\footnote{The HeunC function is
the solution of the Heun Confluent equation which is defined as
%\begin{eqnarray} \label{sp1}
\[
X''(r)-\frac{1+\beta-(\alpha-\beta-\gamma-2)r-r^2\alpha}{r(r-1)}X'(r)-\frac{\alpha(1+\beta)-\gamma-2\eta-(1+\gamma)\beta-r(2\delta+[2+\gamma+\beta])}
{2r(r-1)}X(r)=0\,.
%\end{eqnarray}
\]
The solution of the above differential equation defined $\mathrm{HeunC}(\alpha,\beta,\gamma,\delta,\eta,r)$ for more details, interested readers can check.
The HeunCPrime is the derivative of the Heun Confluent function.}.
Finally the erf function is the error function which is defined by
\begin{eqnarray}
\label{err}
\mathrm{erf}(x)=\frac{2}{\sqrt{\pi}} \int_0^x\e^{-t^2}dt\,.
\end{eqnarray}
Using Eq.~(\ref{ass1}) in the trace equation, i.e., the fourth equation of Eq.~(\ref{feq}) we get $f(r)$ in the form
\begin{align}
\label{ass11}
f(r)=&-\frac{2 \e^{{\frac{3a_1}{2r^2}}}}{r^7}\left\{2r^3\left( 3r^2 \left[ r^2+a_1 \right]{H}_1+2a_1 \left[ r^2+3a_1 \right]{H}_3 \right)
\int\frac{\e^{^{\frac{-3a_1}{2r^2}}}{H} \left[ r^2+13a_1 \right]}{r \left( r^2+a_1 \right) \left[ \left(2a_1{H}_2-3r^2{H} \right) {H}_1-2a_1{H}{H}_3 \right]}dr \right. \nonumber\\
&+8a_1\left( 3r^2{H} - \left[ 3a_1+r^2 \right]{H}_2 \right)\int\frac{\e^{^{\frac{-3a_1}{2r^2}}}{H}_1r^2 \left[ r^2+13a_1 \right]}{\left( r^2+a_1 \right)
\left( 2a_1{H}_2-3r^2{H} \right){H}_1-2a_1{H}{H}_3}dr+4a_1 \left\{ a_3r^3{H}_3+a_2{H}_2 \right\} \left[ r^2+3a_1 \right]\nonumber\\
& \left. +6r^2 \left( a_3r^3 \left[ r^2+3a_1 \right]{H}_1-2a_1a_2{H} \right) + 2r^3 \e^{^{\frac{-3a_1}{2r^2}}} \left[ r^2+7a_1 \right] \right\}\,.
\end{align}
Using Eq.~(\ref{ass11}) in Eq.~(\ref{Ricci}), we get
\begin{align}
\label{sol11}
{R}=&-\frac{2\e^{^{\frac{3a_1}{2r^2}}}}{r^5 \left( r^2+a_1 \right)}\left\{2r^3 \left( 3r^2 \left[ 2r^2+a_1 \right] {H}_1+2a_1 \left[ 2r^2+3a_1 \right]{H}_3 \right)
\int\frac{\e^{-{\frac{3a_1}{2r^2}}}{H} \left[ r^2+13a_1 \right]}{r \left( r^2+a_1 \right) \left[ \left( 2a_1{H}_2 - 3r^2{H} \right){H}_1-2a_1{H}{H}_3 \right]}dr \right. \nonumber\\
&+4a_1 \left( 3r^2{H} - \left[ 3a_1+2r^2 \right] {H}_2 \right) \int\frac{\e^{^{\frac{-3a_1}{2r^2}}}{H}_1r^2 \left[ r^2+13a_1 \right]}
{\left( r^2+a_1 \right) \left( 2a_1{H}_2 - 3r^2{H} \right){H}_1-2a_1{H}{H}_3}dr+2a_1a_3r^3 \left[ 2r^2+3a_1 \right]{H}_3 \nonumber\\
& \left. +2a_1a_2 \left[ 2r^2+3a_1 \right]{H}_2+3a_3r^5 \left[ 2r^2+a_1 \right]{H}_1-6a_1a_2r^2{H}-2r^3 \e^{-{\frac{3a_1}{2r^2}}} \left[ r^2+7a_1 \right] \right\}.
\end{align}
Equations~(\ref{ass1}), (\ref{ass11}), and (\ref{sol11}) show that when $a_1=0$, we get
\begin{eqnarray}
\label{reda}
\alpha(r)=1\,, \quad f(r)=-6a_3, \quad {R}=-12a_3,\quad \mu(r)=\nu(r)=1+\frac{a_2}{r}+a_3r^2\,,\quad \mbox{and} \qquad F(r)=1\,, \quad {\xi}=a_0\,.
\end{eqnarray}
Equation~(\ref{reda}) indicates that when ${F}_1(r)=1$, we get $f({R})={R}$ which gives $\mu(r)=\nu(r)=1+\frac{a_2}{r}+a_3r^2$.
All the above informations guarantees that when $a_1=0$ we get the GR black holes which makes the constant $a_3$ acts as a cosmological
constant\footnote{Note that when $a_1=0$, we get ${H}={H}_1=\mathrm{HeunC}\left(\frac{3}{2},\frac{3}{2},0,\frac{3}{8},\frac{9}{8},0\right)=\mathrm{HeunC}
\left(\frac{3}{2},-\frac{3}{2},0,\frac{3}{8},\frac{9}{8},0\right)=1$, \\
${H}_2=\mathrm{HeunCPrime}\left(\frac{3}{2},\frac{3}{2},0,\frac{3}{8},\frac{9}{8},-\frac{a_1}{r^2}\right)=0$
and ${H}_3=\mathrm{HeunCPrime}\left(\frac{3}{2},-\frac{3}{2},0,\frac{3}{8},\frac{9}{8},-\frac{a_1}{r^2}\right)=-\frac{3}{2}$.}.
Equation~(\ref{reda}) ensures that in higher-order curvature theory we cannot reproduce the ordinary Reissner-Nordstr\"om.
This is because that the charged solution in this class of modified theory is related to the higher order curvature, i.e., depends on $a_1$ and if $a_1=0$,
we will not get the Reissner-Nordstr\"om solution of GR.
%%%%%%%%%%%%%%%%%%%%%%%%%%%%%%%%%%% Section 3 %%%%%%%%%%%%%%%%%%%%%%%%%%%%%%%%%%%%%%%%
\subsection{Physical properties of the black hole (\ref{ass1})}
%%%%%%%%%%%%%%%%%%%%%%%%%%%%%%%%%%%%%%%%%%%%%%%%%%%%%%%%%%%%%%%%%%%%%%%%%%%%%%%%%%%%%%
In this section, we try to understand the physical properties of the black hole (\ref{ass1}).
The asymptote behaviors of the metric potentials, $\mu(r)$ and $\nu(r)$ in Eq.~(\ref{ass1}), by assuming  $a_3=0$, have the form
\begin{align}
\label{mpab}
\mu(r)\approx& 1-\frac{2M}{r}-\frac{13a_1}{2r^2}-\frac{a_1(15M-4a_1)}{10r^3}+\frac{379a_1a_1{}^2}{21r^4}\cdots\,,\nonumber\\
\nu(r)\approx& 1-\frac{2M}{r}-\frac{7a_1}{2r^2}+\frac{a_1(15M+4a_1)}{10r^3}+\frac{64a_1{}^2}{21r^4}+\cdots\,.
\end{align}
In Eq.~(\ref{mpab}), we have put $a_2=-2M$.
Using Eq.~(\ref{mpab}) in Eq.~(\ref{met12}), we get
\begin{align}
\label{metaf}
ds^2\approx & -\left[ 1-\frac{2M}{r}-\frac{13a_1}{2r^2}-\frac{a_1 \left( 15M-4a_1 \right)}{10r^3}+\frac{379a_1a_1{}^2}{21r^4} \right]dt^2
+\frac{dr^2}{ 1-\frac{2M}{r}-\frac{7a_1}{2r^2}+\frac{a_1 \left( 15M+4a_1 \right)}{10r^3}+\frac{64a_1{}^2}{21r^4}}\nonumber\\
& +r^2 \left(d\theta^2+\sin^2d\phi^2 \right)\,.
\end{align}
The line element (\ref{metaf}) is asymptotically approaching a flat spacetime and does not coincide with the Reissner-Nordstr\"om spacetime
due to the contribution of the extra terms that come mainly from the constant parameter $a_1$ whose source is the higher-order curvature terms of $f( R)$ gravitational theory.
Equation~(\ref{metaf}) shows in a clear way that in the higher-order curvature gravity, one can get a spacetime different from the Reissner-Nordstr\"om
and when the constant $a_1=0$, we can recover the Reissner-Nordstr\"om metric.
As one can check easily that when these extra terms equal zero one can smoothly return to the Schwarzschild spacetime.
In conclusion, we can say that in higher-order curvature gravity, we can get a charged spacetime that is different from the Reissner-Nordstr\"om
and cannot reduce to the Reissner-Nordstr\"om in the lower order of $f(R)=R$.

Now we going to assume that the constant $a_3\neq 0$ and get the metric potentials in the form
\begin{eqnarray}
\label{mpab1}
&& \mu(r)\approx \Lambda_\mathrm{eff} r^2+1-\frac{2M}{r}-\frac{13a_1}{2r^2}-\frac{a_1 \left( 15M-4a_1 \right)}{10r^3}+\frac{379a_1a_1{}^2}{21r^4}+\cdots\,,\nonumber\\
&&
\nu(r)\approx \Lambda_\mathrm{eff} \left( r^2+3a_1 \right) + 1-\frac{2M}{r}+\frac{a_1 \left( 9a_1{}a_3-7a_1 \right)}{2r^2}
+ \frac{a_1 \left( 15M+4a_1 \right)}{10r^3} + \frac{a_1{}^2 \left( 128+189 a_1\Lambda_\mathrm{eff} \right)}{42r^4}\cdots\, ,
\end{eqnarray}
where $\Lambda_\mathrm{eff}=a_3$.
Uses of Eq.~(\ref{mpab1}) in (\ref{met12}) gives
\begin{align}
\label{metaf1}
ds^2\approx & -\left[ \Lambda_\mathrm{eff} r^2+1-\frac{2M}{r}-\frac{13a_1}{2r^2}-\frac{a_1 \left( 15M-4a_1 \right)}{10r^3}
+\frac{379a_1a_1{}^2}{21r^4} \right] dt^2 \nonumber\\
& + \left[ \Lambda_\mathrm{eff}(r^2+3a_1)+1-\frac{2M}{r}+\frac{a_1 \left( 9a_1{}a_3-7a \right)}{2r^2}+\frac{a_1 \left( 15M+4a_1 \right)}{10r^3}
+ \frac{a_1{}^2 \left( 128+189 a_1\Lambda_\mathrm{eff} \right)}{42r^4} \right]^{-1}dr^2 \nonumber\\
& +r^2 \left( d\theta^2+\sin^2d\phi^2 \right)\,.
\end{align}
The line element (\ref{metaf}) is asymptotically approaches the AdS/dS spacetime according to the sign of $\Lambda_\mathrm{eff}$ which depends mainly on the constant $a_3$.

Now we are going to use Eq.~(\ref{metaf1}) in Eq.~(\ref{Ricci}) and get
\begin{eqnarray}
\label{R1}
&&{R}(r)\approx -12\Lambda_\mathrm{eff}-\frac{30a_1\Lambda_\mathrm{eff}}{r^2}-\frac{24a_1{}^2\Lambda_\mathrm{eff}}{r^3}
 -\frac{2a_1 \left( 21a_1\Lambda_\mathrm{eff}-2 \right)}{r^3}+\cdots
\Rightarrow r(R)=\frac{2^{1/3} \left[ {{f}_3}{}^{2/3}-5a_1{}^{1/3}\Lambda_\mathrm{eff} \right]}{\sqrt{{R}+12\Lambda_\mathrm{eff}}{{f}_3}{}^{1/3}}\,, \nonumber\\
&&\mbox{where} \quad {f}_3=a_1\Lambda_\mathrm{eff} \left[ \sqrt{2a_1(125\Lambda_\mathrm{eff}+18a_1{R}+216a_1\Lambda_\mathrm{eff})}
 -6\Lambda_\mathrm{eff}\sqrt{{R}+12 \Lambda_\mathrm{eff}} \right]\,.
\end{eqnarray}
We have neglected the other two roots of Eq.~(\ref{R1}) because they are imaginary.
From Eq.~(\ref{R1}), it is clear that when the constant $a_3=0$ we get a non-vanishing value of the Ricci scalar
because of the higher-order curvature.
When $a_1=0$ we get a vanishing value of the Ricci scalar which corresponds to GR black hole.
The asymptote form of the function $f(r)$ given by Eq.~(\ref{ass11}) becomes
\begin{eqnarray}
\label{fR1}
&&f(r)\approx -6\Lambda_\mathrm{eff}-\frac{24a_1a_3}{r^2}-\frac{12a_1{}^2a_3}{r^3}+\frac{a_1(14a_1-45a_1a_3)}{r^4}\cdots\,.
\end{eqnarray}
Using second equation of (\ref{R1}) in (\ref{fR1}), we get
\begin{eqnarray}
\label{fR2}
&&f(R)\approx C_1+C_2R+C_3R^2+C_4R^3+\cdots\,,\nonumber\\
\end{eqnarray}
where $C_i$, $i=1,\cdots , 4$ are lengthy constants whose values depend on $a_1$ and $a_3$.

{ Using Eq.~(\ref{mpab1}), we get the invariants of solution (\ref{ass1}) as}\footnote{ We write the exact form of the Kretschmann
scalar in Appendix A to show that it is divergent at the origin:}
\begin{align}
\label{inv}
R_{\mu \nu \rho \sigma} R^{\mu \nu \rho \sigma}\approx & 24\Lambda_\mathrm{eff}{}^2+\frac{120a_1\Lambda_\mathrm{eff}{}^2}{r^2}
+\frac{96a_1{}^2\Lambda_\mathrm{eff}{}^2}{r^3}-\frac{88a_1\Lambda_\mathrm{eff}}{r^4}
+\frac{384a_1{}^2\Lambda_\mathrm{eff}{}^2}{r^5}\nonumber\\
& +\frac{4a_1\Lambda_\mathrm{eff} \left( 36a_1{}^3\Lambda_\mathrm{eff}+202a_1-12a_1a_2+288a_1{}^2\Lambda_\mathrm{eff} \right)+12a_2{}^2}{r^6}\cdots\,, \nonumber\\
R_{\mu \nu } R^{\mu \nu }\approx & 36\Lambda_\mathrm{eff}{}^2+\frac{180a_1\Lambda_\mathrm{eff}{}^2}{r^2}+\frac{144a_1{}^2\Lambda_\mathrm{eff}{}^2}{r^3}+\frac{4a_1\Lambda_\mathrm{eff}
\left( 126a_1\Lambda_\mathrm{eff}-6 \right)}{r^4}
+\frac{72a_1\Lambda_\mathrm{eff} \left(2a_1+60a_1{}^2\Lambda_\mathrm{eff}-5a_2 \right)}{5r^5} \nonumber\\
&+\frac{12a_1\Lambda_\mathrm{eff} \left( 12a_1{}^3\Lambda_\mathrm{eff}+59a_1+84a_1{}^2\Lambda_\mathrm{eff} \right)}{r^6}
+\frac{72a_1\Lambda_\mathrm{eff} \left( 245a_1{}^3\Lambda_\mathrm{eff}+171a_1{}^2-98a_1a_2 \right)}{49r^7}\nonumber\\
& +\frac{a_1{}^2 \left( 10885+40725a_1\Lambda_\mathrm{eff}+1512a_1{}^2\Lambda_\mathrm{eff}-2160a_1a_2\Lambda_\mathrm{eff}+56070a_1\Lambda_\mathrm{eff}
+35280a_1{}^2\Lambda_\mathrm{eff} \right)}{35r^8}\cdots\,, \nonumber\\
R{{\approx}} & -12\Lambda_\mathrm{eff}-\frac{30a_1\Lambda_\mathrm{eff}}{r^2}-\frac{24a_1{}^2\Lambda_\mathrm{eff}}{r^3}
 -\frac{2a_1 \left( 21a_1\Lambda_\mathrm{eff}-2 \right)}{r^3} \cdots\,.
\end{align}
Here $\left(R_{\mu \nu \rho \sigma} R^{\mu \nu \rho \sigma}, R_{\mu \nu} R^{\mu \nu}, R \right)$ represent the Kretschmann
scalar, the Ricci tensor square, the Ricci scalar, respectively and all of them have a true singularity at $r=0$.
It is important to note that the constant $a_1$ is the source of the deviation of the above results of charged solution from the Reissner-Nordstr\"om spacetime
of GR whose invariants have the form $\left( R_{\mu \nu \rho \sigma} R^{\mu \nu \rho \sigma}, R_{\mu \nu} R^{\mu \nu}, R \right)
= \left( 24\Lambda^2+\frac{48M^2}{r^6},36\Lambda^2+\frac{4q^4}{r^8},12\Lambda \right)$.
Equation~(\ref{inv}) indicates that the leading term of the invariants $\left( R_{\mu \nu \rho \sigma} R^{\mu \nu \rho \sigma},R_{\mu \nu} R^{\mu \nu},R \right)$
is $\left( \frac{1}{r^\mathrm{2}},\frac{1}{r^\mathrm{2}},\frac{1}{r^\mathrm{2}} \right)$ which is different from the Reissner-Nordstr\"om black hole
whose leading terms of the Kretschmann and the Ricci tensor squared are $\left( \frac{1}{r^\mathrm{6}},\frac{1}{r^\mathrm{8}} \right)$.
Therefore, Eq.~(\ref{inv}) indicates that the singularity of the Kretschmann scalar and the Ricci tensor squared are much softer than those of Reissner-Nordstr\"om black hole of GR.
%%%%%%%%%%%%%%%%%%%%%%%%%%%% Section 7 %%%%%%%%%%%%%%%%%%%%%%%%%%%%%
\subsection{Stability of the black holes using geodesic deviation}\label{S6336b}
%%%%%%%%%%%%%%%%%%%%%%%%%%%%%%%%%%%%%%%%%%%%%%%%%%%%%%%%%%%%%%%%%%%%
The geodesic equations of a test particle in the gravitational field are given by
\begin{equation}
\label{ge3}
\frac{d^2 x^\alpha}{d\lambda^2} + \left\{ \begin{array}{c} \alpha \\ \beta \rho \end{array} \right\}
\frac{d x^\beta}{d\lambda} \frac{d x^\rho}{d\lambda}=0\, ,
\end{equation}
where $s$ represents the affine connection parameter.
The geodesic deviation equations have the form \cite{Nashed:2003ee}
\begin{equation}
\label{ged333}
\frac{d^2 \epsilon^\sigma}{d\lambda^2} + 2 \left\{ \begin{array}{c} \sigma \\ \mu \nu \end{array} \right\}
\frac{d x^\mu}{d\lambda} \frac{d \epsilon^\nu}{d\lambda} + \left\{ \begin{array}{c} \sigma \\ \mu \nu \end{array} \right\}_{,\ \rho}
\frac{d x^\mu}{d\lambda} \frac{d x^\nu}{d\lambda}\epsilon^\rho=0\, ,
 \end{equation}
where $\epsilon^\rho$ is the 4-vector deviation. Introducing (\ref{ge3}) and (\ref{ged333}) into (\ref{metaf1}), we get
\begin{equation}
\frac{d^2 t}{d\lambda^2}=0\, , \qquad \frac{1}{2} \mu'(r)\left( \frac{d t}{d\lambda}\right)^2 - r \left( \frac{d \phi}{d\lambda}\right)^2=0\, , \qquad
\frac{d^2 \theta}{d\lambda^2}=0\, ,\qquad \frac{d^2 \phi}{d\lambda^2}=0\, ,
\end{equation}
and for the geodesic deviation, the black hole spacetime (\ref{metaf1}) gives
\begin{eqnarray}
\label{ged11}
&& \frac{d^2 \epsilon^1}{d\lambda^2} + \nu(r)\mu'(r) \frac{dt}{d\lambda} \frac{d \epsilon^0}{d\lambda}
 -2r \nu(r) \frac{d \phi}{d\lambda} \frac{d \epsilon^3}{d\lambda} + \left[ \frac{1}{2}\left(\mu'(r)\nu'(r)+\nu(r) \mu''(r)
\right)\left( \frac{dt}{d\lambda}\right)^2 - \left(\nu(r)+r\nu'(r) \right) \left( \frac{d\phi}{d\lambda}\right)^2 \right]\epsilon^1=0\, , \nonumber\\
&& \frac{d^2 \epsilon^0}{d\lambda^2} + \frac{\nu'(r)}{\nu(r)} \frac{dt}{d\lambda} \frac{d \zeta^1}{d\lambda}=0\, ,\qquad
\frac{d^2 \epsilon^2}{d\lambda^2} + \left( \frac{d\phi}{d\lambda}\right)^2 \epsilon^2=0\, , \qquad
\frac{d^2 \epsilon^3}{d\lambda^2} + \frac{2}{r} \frac{d\phi}{d\lambda} \frac{d \epsilon^1}{d\tau}=0\, ,
\end{eqnarray}
where $\mu(r)$ and $\nu(r)$ are defined in Eq.~(\ref{metaf1}) and $'$ means derivative w.r.t. the radial coordinate $r$.
Using the circular orbit
\begin{equation}
\theta= \frac{\pi}{2}\, , \qquad
\frac{d\theta}{d\lambda}=0\, , \qquad \frac{d r}{d\lambda}=0\, ,
\end{equation}
we get
\begin{equation}
\left( \frac{d\phi}{d\lambda}\right)^2 = \frac{\mu'(r)}{r \left[ 2\mu(r)-r\mu'(r) \right]}\, , \qquad
\left( \frac{dt}{d\lambda} \right)^2 = \frac{2}{2\mu(r)-r\mu'(r)}\, .
\end{equation}

\begin{figure}[ht]
\centering
\subfigure[~Stability of the black hole for the case $a_3= 0$]{\label{fig:1a}\includegraphics[scale=0.3]{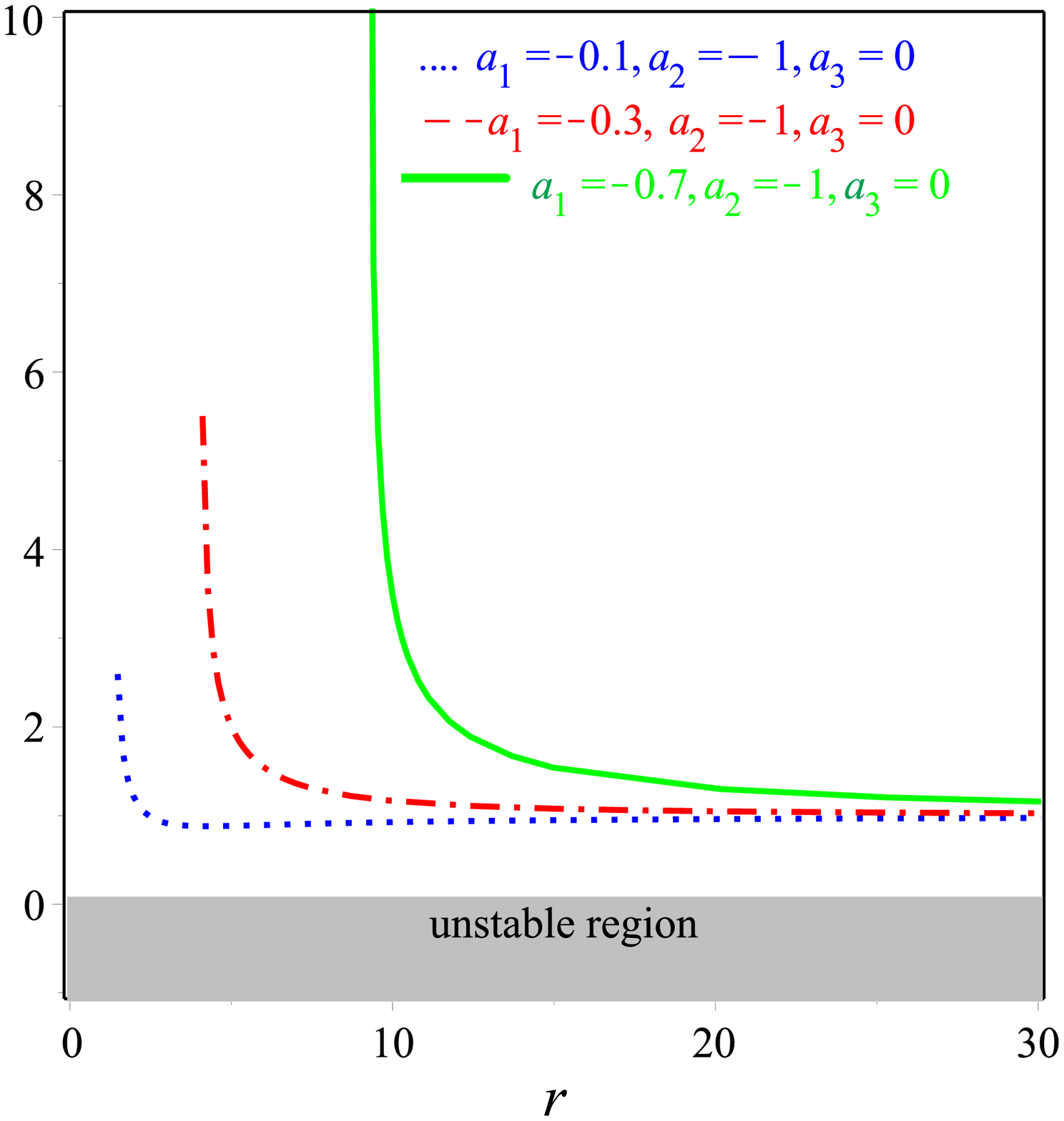}}\hspace{0.2cm}
\subfigure[~Stability of the black hole for the case $a_3\neq 0$]{\label{fig:1b}\includegraphics[scale=0.3]{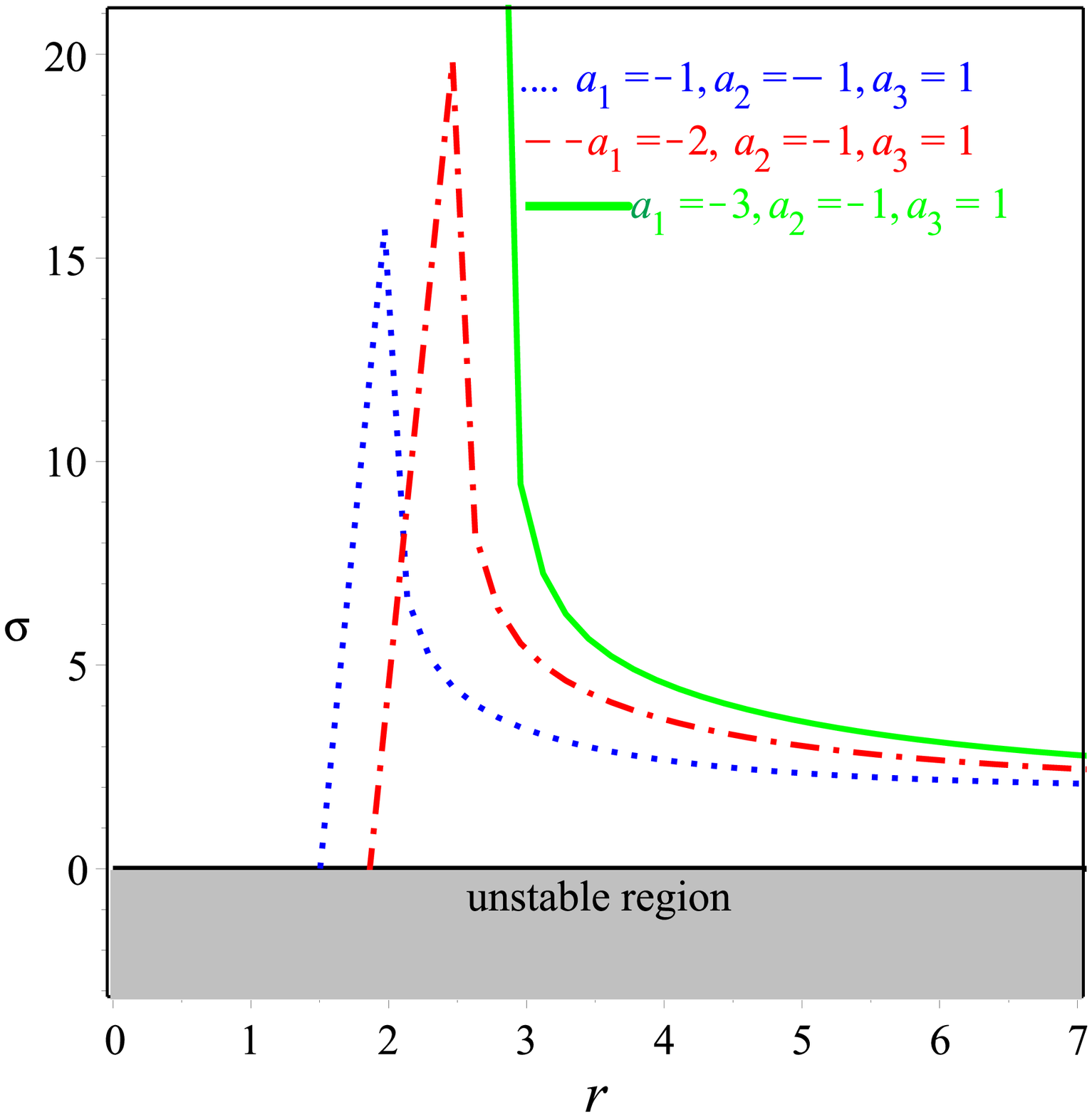}}
\caption{ {Schematic plot of Eq.~(\ref{stab1}), namely $\sigma^2$ versus the coordinate $r$ using the two black holes (\ref{mpab}) and (\ref{mpab1}).}}
\label{Fig:1}
\end{figure}

Equations~(\ref{ged11}) can be rewritten as
\begin{eqnarray}
\label{ged22}
&&
\frac{d^2 \epsilon^1}{d\phi^2} + \mu(r)\mu'(r) \frac{dt}{d\phi} \frac{d \epsilon^0}{d\phi} - 2r \nu(r) \frac{d \epsilon^3}{d\phi} + \left[ \frac{1}{2} \left[\mu'^2(r)+\mu(r) \mu''(r)
\right] \left( \frac{dt}{d\phi} \right)^2-\left[\mu(r)+r\mu'(r) \right] \right]\zeta^1=0\, , \nonumber\\
&& \frac{d^2 \epsilon^2}{d\phi^2} + \epsilon^2=0\, , \qquad
\frac{d^2 \epsilon^0}{d\phi^2} + \frac{\mu'(r)}{\mu(r)} \frac{dt}{d\phi} \frac{d \epsilon^1}{d\phi}=0\, ,\qquad
\frac{d^2 \epsilon^3}{d\phi^2} + \frac{2}{r} \frac{d \epsilon^1}{d\phi}=0\, .
\end{eqnarray}
The second equation of (\ref{ged22}) corresponds to a simple harmonic motion, which means that there is stability on the plane $\theta=\pi/2$.
Assuming the remaining equations of (\ref{ged22}) to have solutions in the form
\begin{equation}
\label{ged33}
\epsilon^0 = \zeta_1 \e^{i \sigma \varphi}\, , \qquad
\epsilon^1= \zeta_2\e^{i \sigma \varphi}, \qquad \mbox{and} \qquad \epsilon^3 = \zeta_3 \e^{i \sigma \varphi}\, ,
\end{equation}
where $\zeta_1$, $\zeta_2$, and $\zeta_3$ are constants and $\varphi$ is an unknown variable.
Using Eq.~(\ref{ged33}) into (\ref{ged22}), one can get the stability condition for a static spherically symmetric charged black hole in the form
\begin{equation}
\label{con111}
\frac{3\mu \nu \mu'-\sigma^2\mu \mu'-2r\nu\mu'^{2}+r\mu \nu \mu'' }{\mu \nu'}>0\, .
\end{equation}
Equation~(\ref{con111}) can has the following solution
\begin{equation}
\label{stab1}
\sigma^2= \frac{3\mu \nu \mu''-2r\nu\mu'^{2}+r\mu \nu \mu'' }{\mu^2 \nu'^2}>0\, .
\end{equation}
We depicted Eq.~(\ref{stab1}) in Fig.~\ref{Fig:1} for particular values of the model.
The case $\Lambda_\mathrm{eff}=a_3= 0$ is drawn in Fig.~\ref{Fig:1}~\subref{fig:1a} and the case $\Lambda_\mathrm{eff}=a_3\neq 0$ is drawn
in Fig.~\ref{Fig:1}~\subref{fig:1b}.
These two figures exhibit the regions where the black holes are stable and not stable.
%%%%%%%%%%%%%%%%%%%%%%%%%%%%%%%%%%% Section 4 %%%%%%%%%%%%%%%%%%%%%%%%%%%%%%%%%%%%%%%%
\section{Thermodynamics of the black hole}\label{S4}
\begin{figure}
\centering
\subfigure[~The metric potentials of the black hole (\ref{mpab})]{\label{fig:met}\includegraphics[scale=0.23]{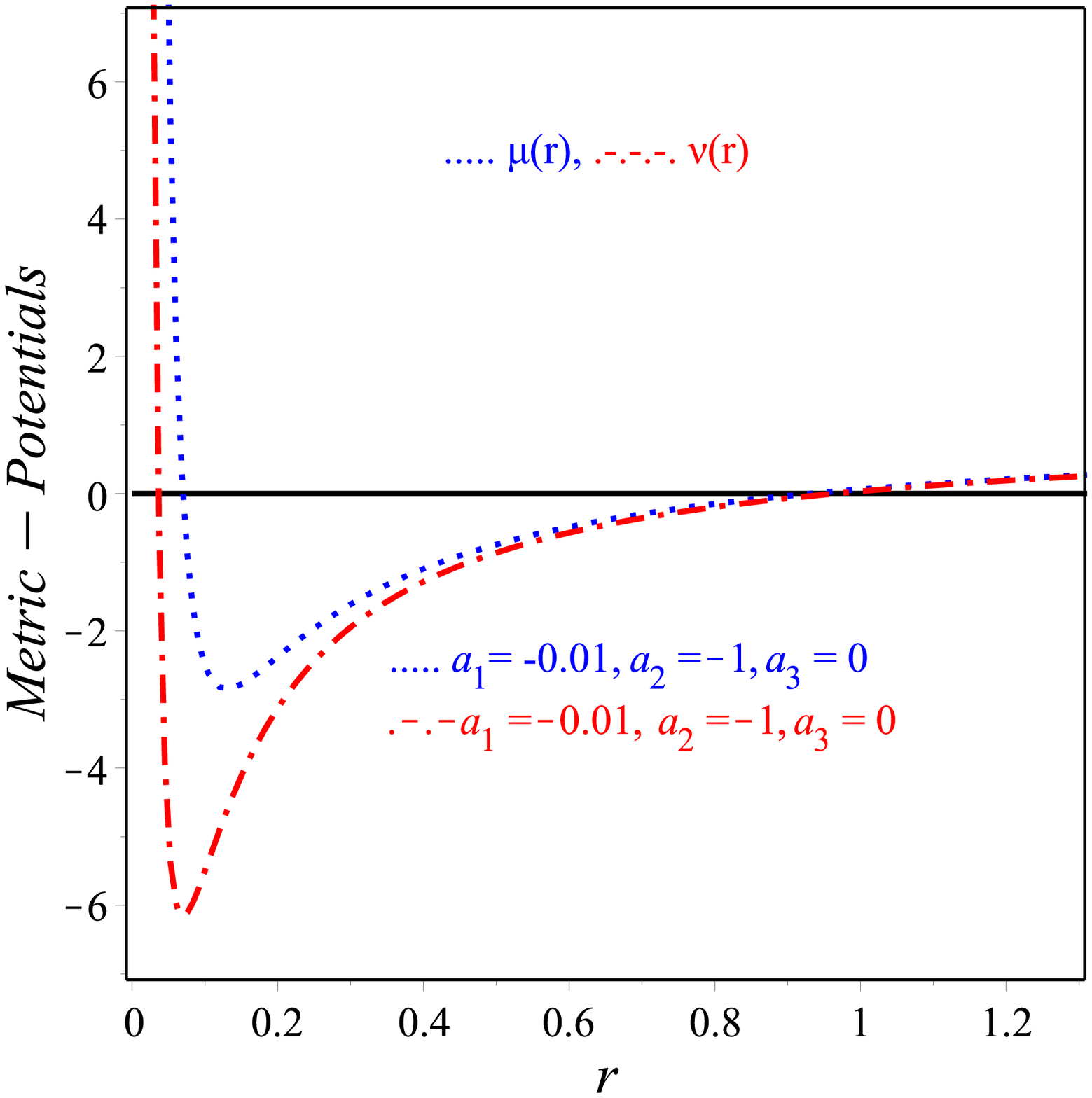}}
\subfigure[~The horizons of the black hole (\ref{mpab}) of the metric potential $\mu$]{\label{fig:metrd}\includegraphics[scale=0.23]{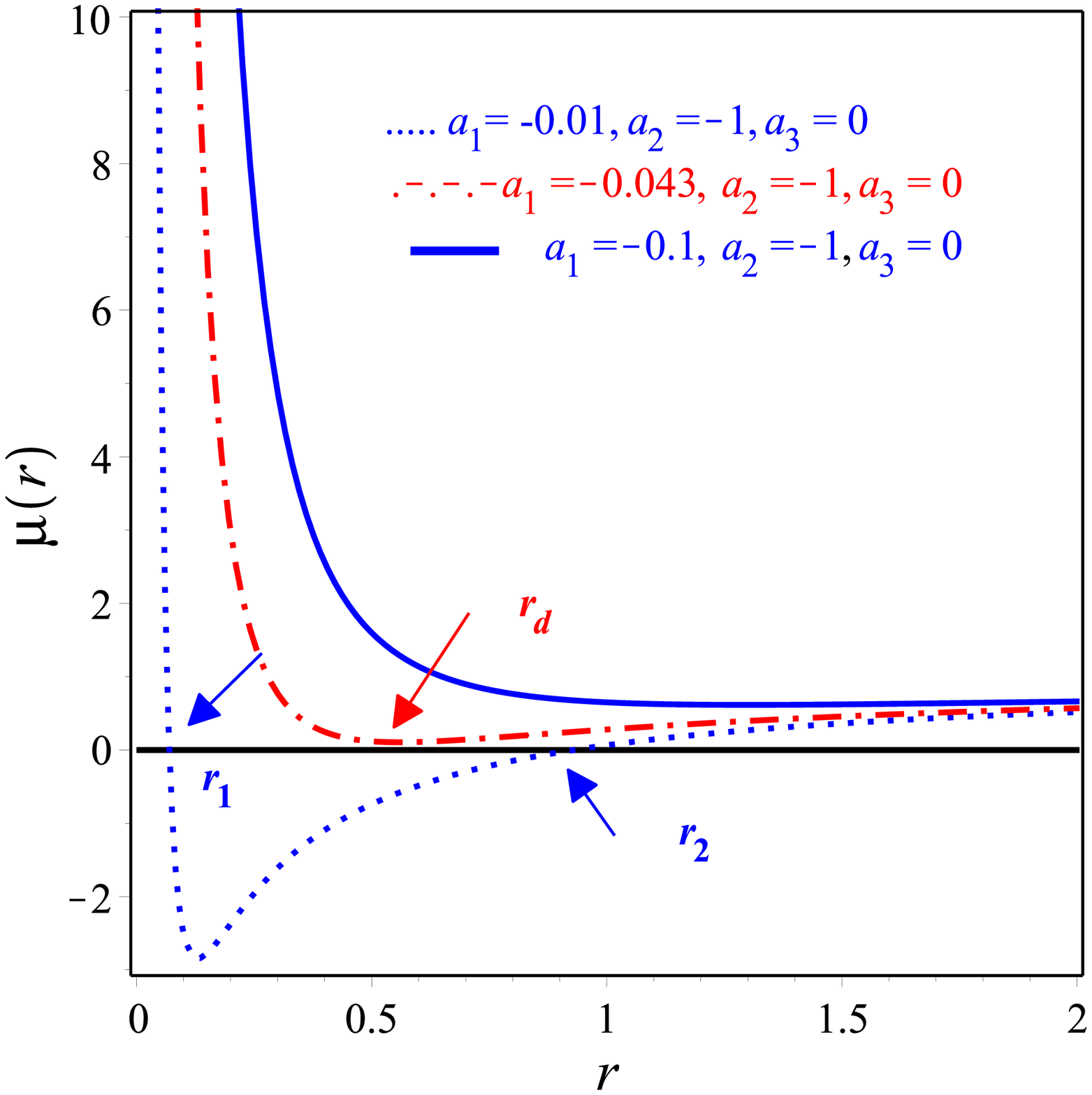}}
\subfigure[~Hawking temperature of the black hole (\ref{mpab})]{\label{fig:temp}\includegraphics[scale=0.23]{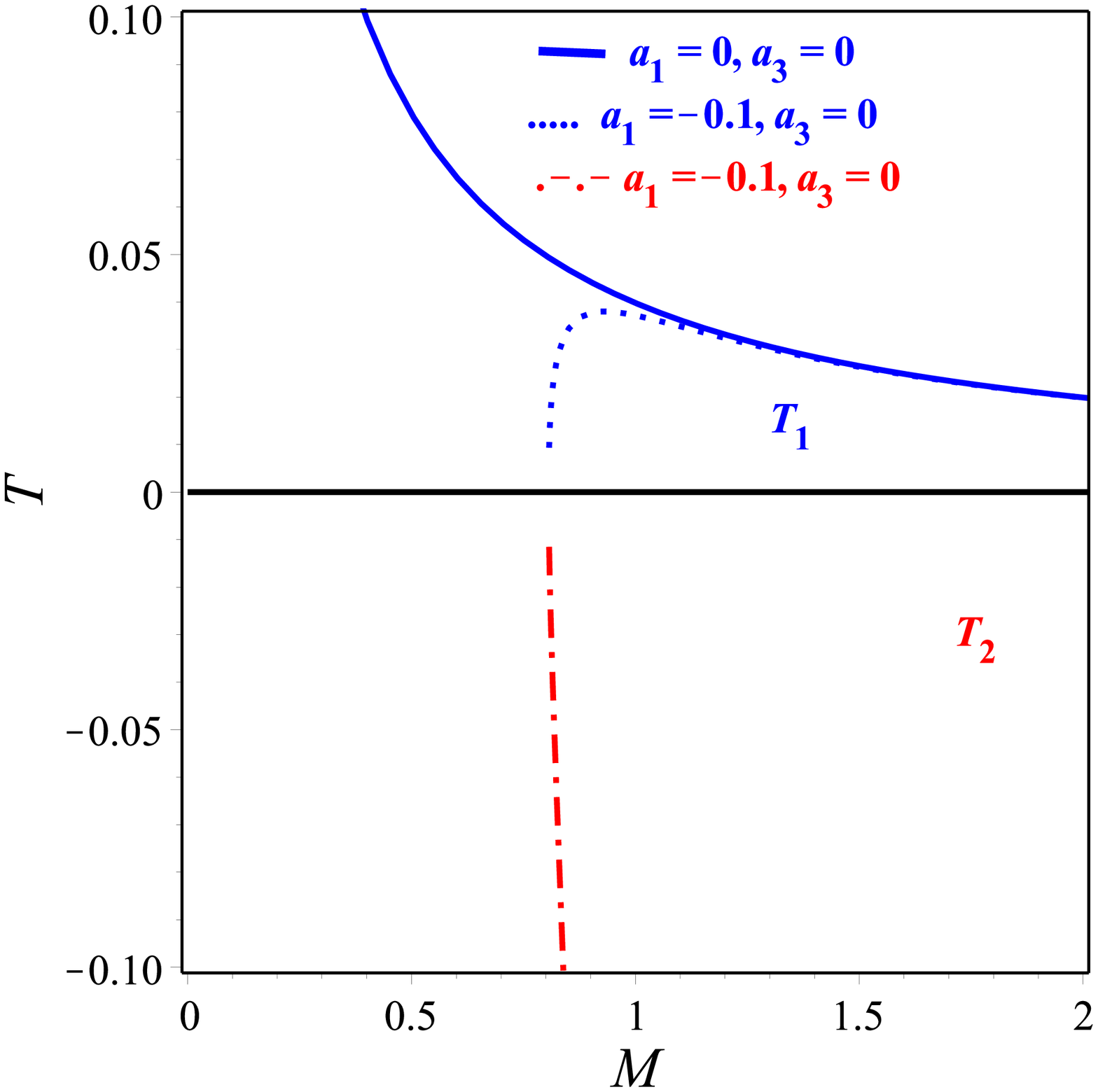}}
\subfigure[~The entropy of the black hole (\ref{mpab})]{\label{fig:ent}\includegraphics[scale=0.23]{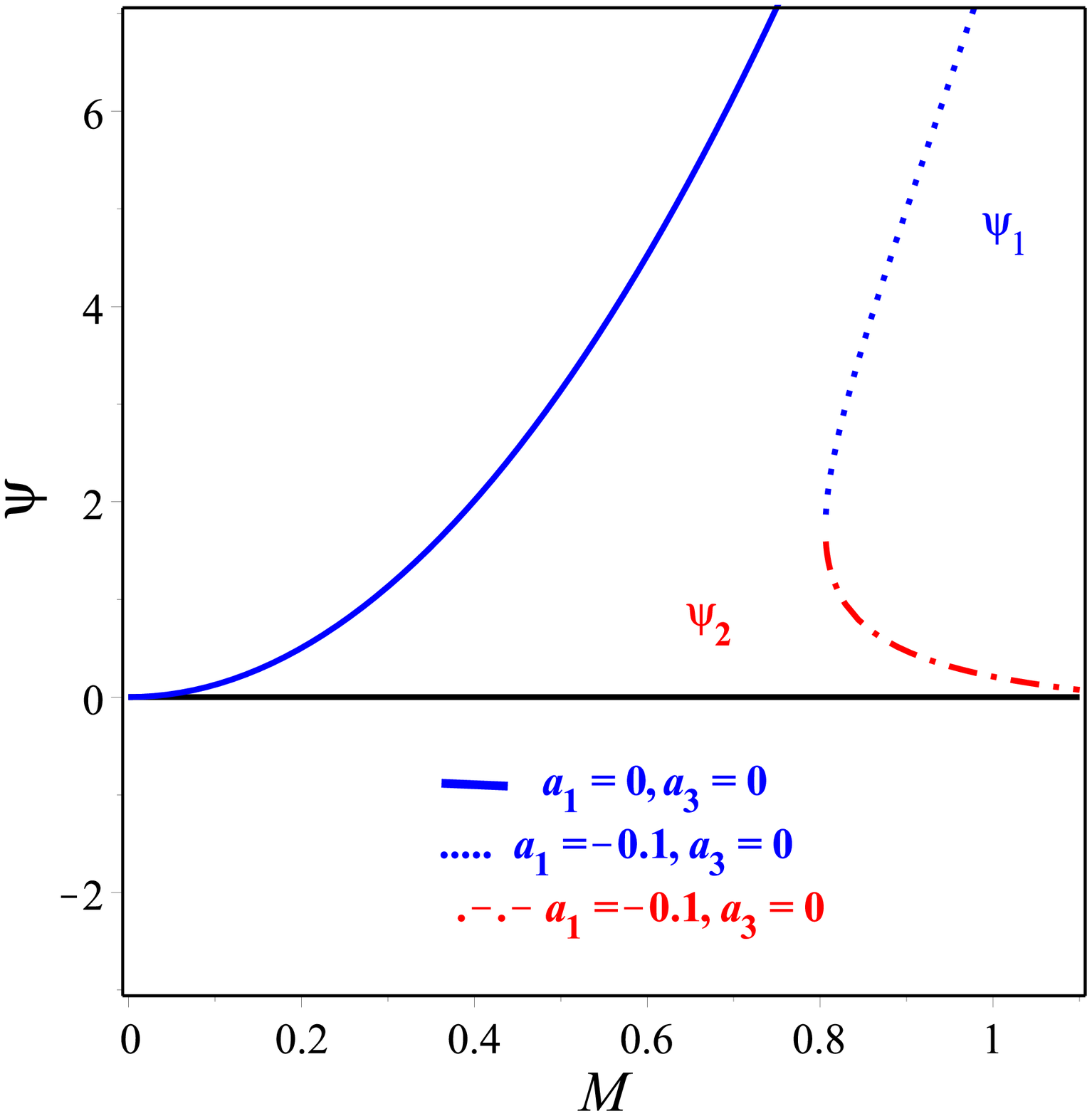}}
\subfigure[~The quasi-local energy of the black hole (\ref{mpab})]{\label{fig:enr}\includegraphics[scale=0.23]{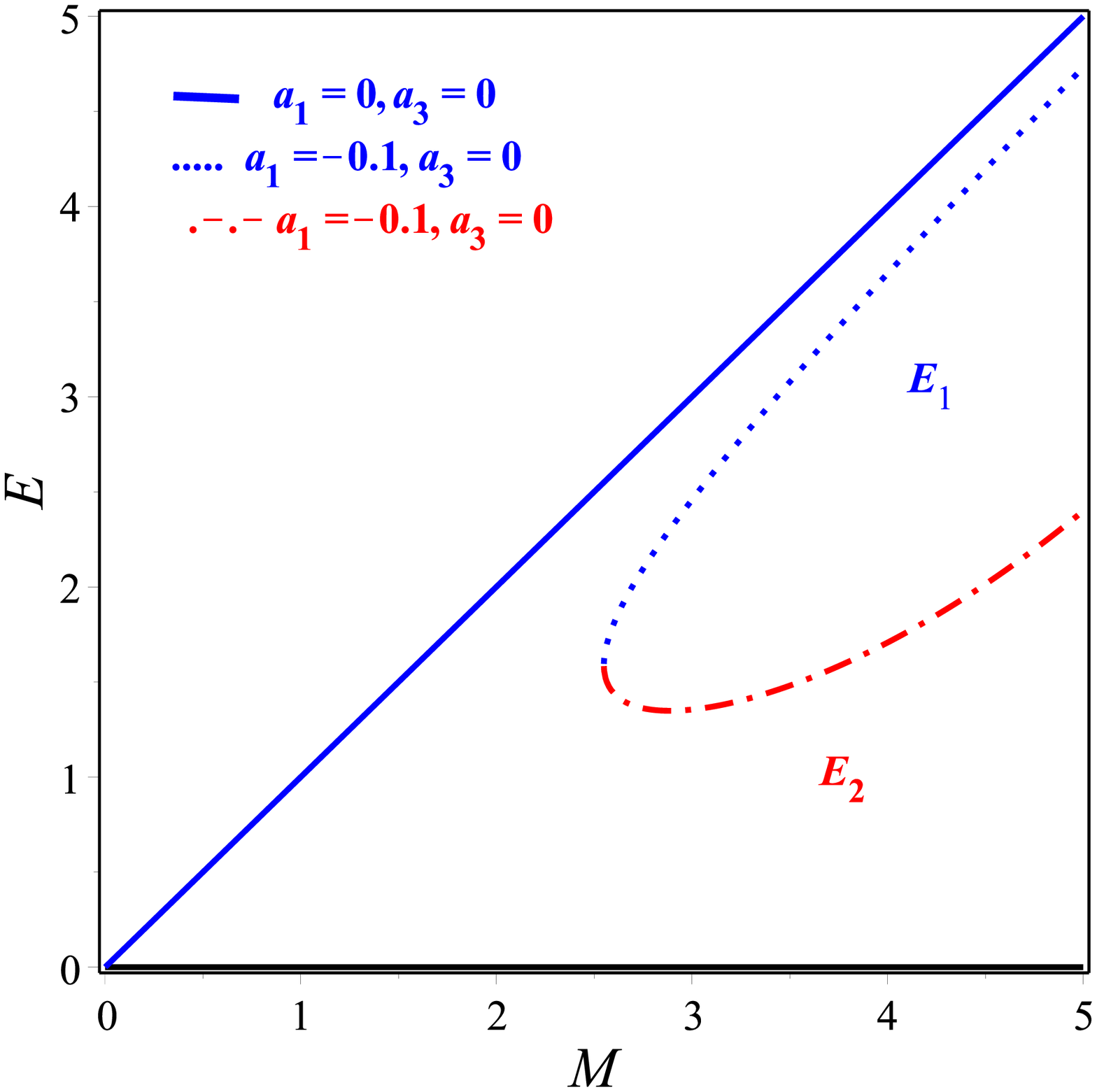}}
\subfigure[~The free energy of the black hole (\ref{mpab})]{\label{fig:gib}\includegraphics[scale=0.23]{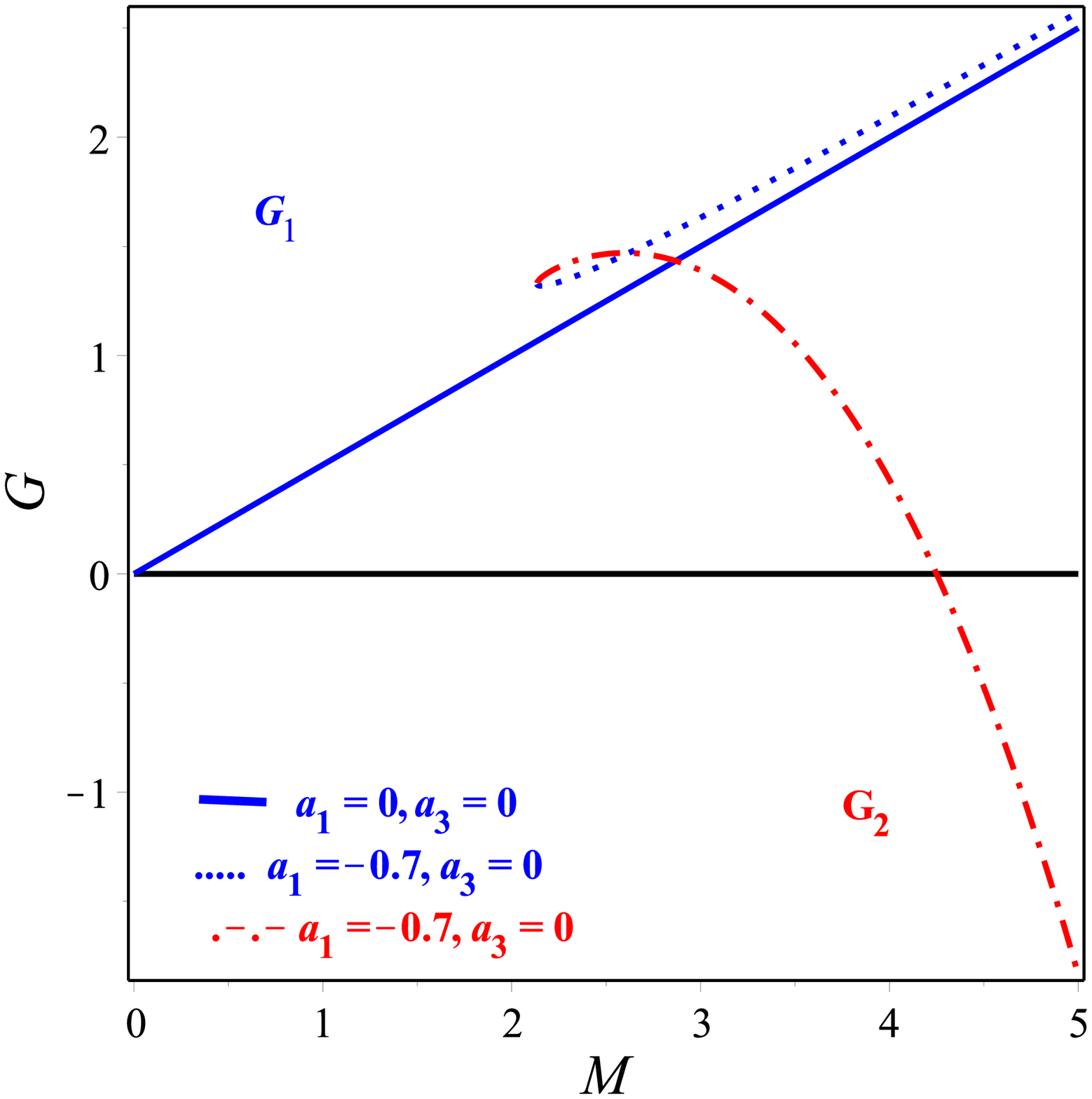}}
\caption[figtopcap]{Schematic plots of thermodynamical quantities of the black hole solution (\ref{mpab}): \subref{fig:met} shows the behavior
of the metric potentials $\mu$ and $\nu$; \subref{fig:metrd} Typical behaviour of the horizons of the metric potential $\mu(r)$ given
by (\ref{mpab}) ; \subref{fig:temp} typical behaviour of the horizon temperature, (\ref{T1}), which shows that $T_1$ has a positive decreasing value
while $T_2$ has a negative increasing value; \subref{fig:ent} typical behaviour of the horizon entropy, (\ref{S1}), which shows that $\psi_{(1,2)}$ always
has positive values, $ \psi_{1}$ increases with $M$ wile $\psi_{2}$ has positive decreasing value; \subref{fig:enr} typical behaviour
of the quasi-local energy, (\ref{E1}), that indicates that $E_{(1,2)}$ has positive increasing values and also $E_1>E_2$; \subref{fig:gib} typical
behaviour of the horizon Gibbs free energy which shows that $G_1$ is positive while $G_2$ starts with positive value and as $M$ increase it becomes negative.}
\label{Fig:2}
\end{figure}

%%%%%%%%%%%%%%%%%%%%%%%%%%%%%%%%%%%%%%%%%%%%%%%%%%%%%%%%%%%%%%%%%%%%%%%%%%%%%%%%%%%%%%
Here we are going to study the thermodynamical quantities of the black holes (\ref{mpab}) and
(\ref{mpab1})\footnote{In this study we will not touch the black holes (\ref{mpab}) and (\ref{mpab1}) because they have four and eight roots
and it is not easy to derive such roots explicitly.}.
The temperature of a black hole is defined as \cite{PhysRevD.86.024013,Hendi:2010gq,PhysRevD.81.084040,Wang:2018xhw}
\begin{equation}
\label{temp}
T_{(1,2)} = \frac{\xi_{\pm}}{2\pi}=\frac{r_{(1,2)}-r_{(2,1)}}{4\pi r_{(1,2)}{}^2}\,,
\end{equation}
where $r_{(1,2)}$ are the inner and outer horizons of the spacetime.
The Hawking entropy of the horizons is defined as
\begin{equation}
\label{ent}
\psi_{(1,2)} =\frac{1}{4}A_{(1,2)}\,f_{R}\,,
\end{equation}
where $A_{(1,2)}$ is the area of the horizons.
The quasi-local energy is figured out as \cite{PhysRevD.84.023515,PhysRevD.86.024013,Hendi:2010gq,PhysRevD.81.084040,Zheng:2018fyn}
\begin{equation}
\label{en}
E \left( r_{(1,2)} \right)=\frac{1}{4}\displaystyle{\int } \left[ 2f_{{R}}~r_{(1,2)}+r_{(1,2)}{}^2 \left\{ f(R(r_{(1,2)}))-R(r_{(1,2)})f_R~r_{(1,2)} \right\} \right] dr_{(1,2)}\, .
\end{equation}
Finally, the Gibbs free energy in the grand canonical ensemble, is figured out as \cite{Kim:2012cma}
\begin{equation}
\label{enr}
G(r_{(1,2)})=E(r_{(1,2)})-T(r_{(1,2)})S(r_{(1,2)})\, .%+P(r_+)V(r_+),
\end{equation}
The black hole (\ref{mpab}) is characterized by the mass $M$ and the parameter $a_1$ and when the parameter $a_1$ is vanishing we get
the Schwarzschild black hole\footnote{This means that in higher-order curvature theory the charged black hole is not similar to
the Reissner-Nordstr\"om black hole due to the contribution of higher-order curvature terms which is expected.
However, in the lower order, $f(R)=R$, we expect to get the Reissner-Nordstr\"om black hole but this not happen as is clear from Eq.~(\ref{mpab}).
This means that in modified gravitational theory, $f( R)$, we get a new charged black hole which cannot reduce to the Reissner-Nordstr\"om black hole
in the lower order.} which corresponds to GR.
To find the horizons of the black hole, (\ref{mpab}), we put $\mu(r)=0$ and get four lengthy real positive roots.
%These real roots have the form
%\begin{eqnarray}\label{r1}
%&&r_\pm=\frac{[14910M^{4/3}+\sqrt{710}X_1{}^2]X_1\pm\sqrt{710}\sqrt{Y_1X_1{}^2+7.008182433\times10^7M(5M^3-5Mb^2-8b^4)}}{29820M^{1/3}X_1}\,,\nonumber\\
%
%&& r_-=\frac{[14910M^{4/3}+\sqrt{710}X_1{}^2]X_1-\sqrt{710}\sqrt{Y_1X_1{}^2+7.008182433\times10^7(5M^3-5Mq^2-8b^4)}}{29820M^{1/3}X_1}\,,
%\end{eqnarray}
%where %$X_1=\Bigg[993.4163240(112M+1625)b^{8/3}+313110M^{8/3}+1043700b^2M^{2/3}+9.398818978\times10^5b^{4/3}M^{4/3}+4.172348561\times10^5b^{2/3}M^2\Bigg]^{1/4}$
% and $Y_1=\Bigg[-993.4163240(112M+1625)b^{8/3}+626220M^{8/3}+2087400b^2M^{2/3}-9.398818978\times10^5b^{4/3}M^{4/3}-4.172348561\times10^5b^{2/3}M^2\Bigg]$. Equation (\ref{r1}),
%$r_\pm$, put the following constrain to have a real value \begin{eqnarray}\label{cons}
%Y_1X_1{}^2+7.008182433\times10^7(5M^3-5Mq^2-8b^4)>0.\end{eqnarray}
The metric potentials of the black hole (\ref{mpab}) are drawn in Fig.~\ref{Fig:2}~\subref{fig:met}.
 From Fig.~\ref{Fig:2}~\subref{fig:met}, we can easy see the two positive horizons of the metric potentials $\mu(r)$.
It is easy to check that the degenerate horizon for the metric potential $\mu(r)$ is happened for a specific values of $(a_1,M,r)\equiv(-0.043,0.5,0.5)$, respectively,
which correspond to the Nariai black hole.
The degenerate behavior is shown in Fig.~\ref{Fig:2}~\subref{fig:metrd}.
Fig.~\ref{Fig:2}~\subref{fig:metrd} shows that the horizon $r_1$ increases as $a_1$ increases while $r_2$ decreases as $a_1$ decreases.
As we observe from Fig.~\ref{Fig:2}~\subref{fig:metrd} that as $a_1$ increases and $M$ decrease i,e, $a_1 > M$, we enter in a parameter region
where there is no horizon, and thus the central singularity is a naked singularity.
Using Eq.~(\ref{temp}), we get the Hawking temperature in the form
\begin{eqnarray}
\label{T1}
T_{(1,2)}= \pm\frac{\sqrt{4M^2+26a_1}}{\pi \left( 2M\pm\sqrt{4M^2+26a_1} \right)^2}\,.
\end{eqnarray}
The behavior of the Hawking temperature given by Eq.~(\ref{T1}) is drawn in Fig.~\ref{Fig:2}~\subref{fig:temp} which shows that $T_{1}>T_{2}$.
As Fig.~\ref{Fig:2}~\subref{fig:temp} shows that the $T_1$ has a decreasing positive temperature while $T_2$ has an increasing negative temperature.

Using Eq.~(\ref{ent}) we get the entropy of black hole (\ref{mpab}) in the form
\begin{eqnarray}
\label{S1}
\psi_{(1,2)}= \frac{\pi \left( 4M^2\pm2M\sqrt{4M^2+26a_1}+15a_1 \right)}{2}\,.
\end{eqnarray}
The behavior of the entropy is given in Fig.~\ref{Fig:1}~\subref{fig:ent} which shows an increasing value for $\psi_1$ and decreasing value for $\psi_2$.
Using Eq.~(\ref{en}) we calculate the quasi-local energy and get
\begin{eqnarray}
\label{E1}
E_{(1,2)}=\frac{\pm6M \left( 4M^2-3a_1 \right)\sqrt{4M^2+26a_1}+96M^4+588a_1M^2\pm6M \left( 4M^2+26a_1 \right)^{3/2}+398a_1{}^2}{3 \left(2M\pm\sqrt{4M^2+26a_1} \right)^3}\,.
\end{eqnarray}
The behavior of the quasi-local energies are shown in Fig.~\ref{Fig:1}~\subref{fig:enr} which also shows positive increasing values for $E_{(1,2)}$ and also shows $E_1>E_2$.
Finally, we use Eqs.~(\ref{T1}), (\ref{S1}) and (\ref{E1}) in Eq.~(\ref{enr}) to calculate the Gibbs free energies.
The behavior of these free energies are shown in Fig.~\ref{Fig:1}~\subref{fig:gib} which shows positive increasing value for $G_1$ and $G_2$ start
with positive decreasing value then it becomes negative value.
%All the above results of the black hole (\ref{mpab}) are physically satisfactory except the result of $-ve$ temperature that goes below absolute zero forming an %ultracold black hole. As noted
%earlier
%by Davies \cite{Davies:1978mf} that there is no obvious reasons from thermodynamics prevent a black hole
%temperature to go below absolute zero and turn it to naked singularity. In fact this is the
%case presented in Fig. \ref{Fig:2} \subref{fig:Temp}.
 %\begin{figure}
%\centering
%\subfigure[~Hawking temperature of black hole (\ref{mpab})]{\label{fig:Temp}\includegraphics[scale=0.4]{JMTAATHNATYASD_temp.eps}}
%\subfigure[~Entropy of black hole (\ref{mpab})]{\label{fig:ent}\includegraphics[scale=0.4]{JMTAATHNATYASD_entropy.eps}}
%
%\caption[figtopcap]{\small{{Plot of the horizons given by Eq.~(\ref{r1}) using $b=0.1$ which is consistent with the constrains (\ref{cons}).}}}
%\label{Fig:2}
%\end{figure}
%\begin{figure}
%\centering
%\subfigure[~Entropy of black hole (\ref{mpab})]{\label{fig:ent}\includegraphics[scale=0.4]{JMTAATHNATYASD_entropy.eps}}
%\subfigure[~Quasi local energy of black hole (\ref{mpab})]{\label{fig:Enr}\includegraphics[scale=0.4]{JMTAATHNATYASD_enr.eps}}
%\subfigure[~Gibbs energy of black hole (\ref{mpab})]{\label{fig:gib}\includegraphics[scale=0.4]{JMTAATHNATYASD_Gib.eps}}
%
%\caption[figtopcap]{\small{{Plot of the horizons given by Eq.~(\ref{r1}) using $b=0.1$ which is consistent with the constrains (\ref{cons}).}}}
%\label{Fig:3}
%\end{figure}

%%%%%%%%%%%%%%%%%%%%%%%%%%%%%%%%%%%% Section 3 %%%%%%%%%%%%%%%%%%%%%%%%%%%%%%%%%%%%%%%%
\subsection{Thermodynamics of the black hole (\ref{mpab1}) that has asymptote flat AdS/dS}
\begin{figure}
\centering
\subfigure[~The metric potential of black hole (\ref{mpab1})]{\label{fig:metl}\includegraphics[scale=0.23]{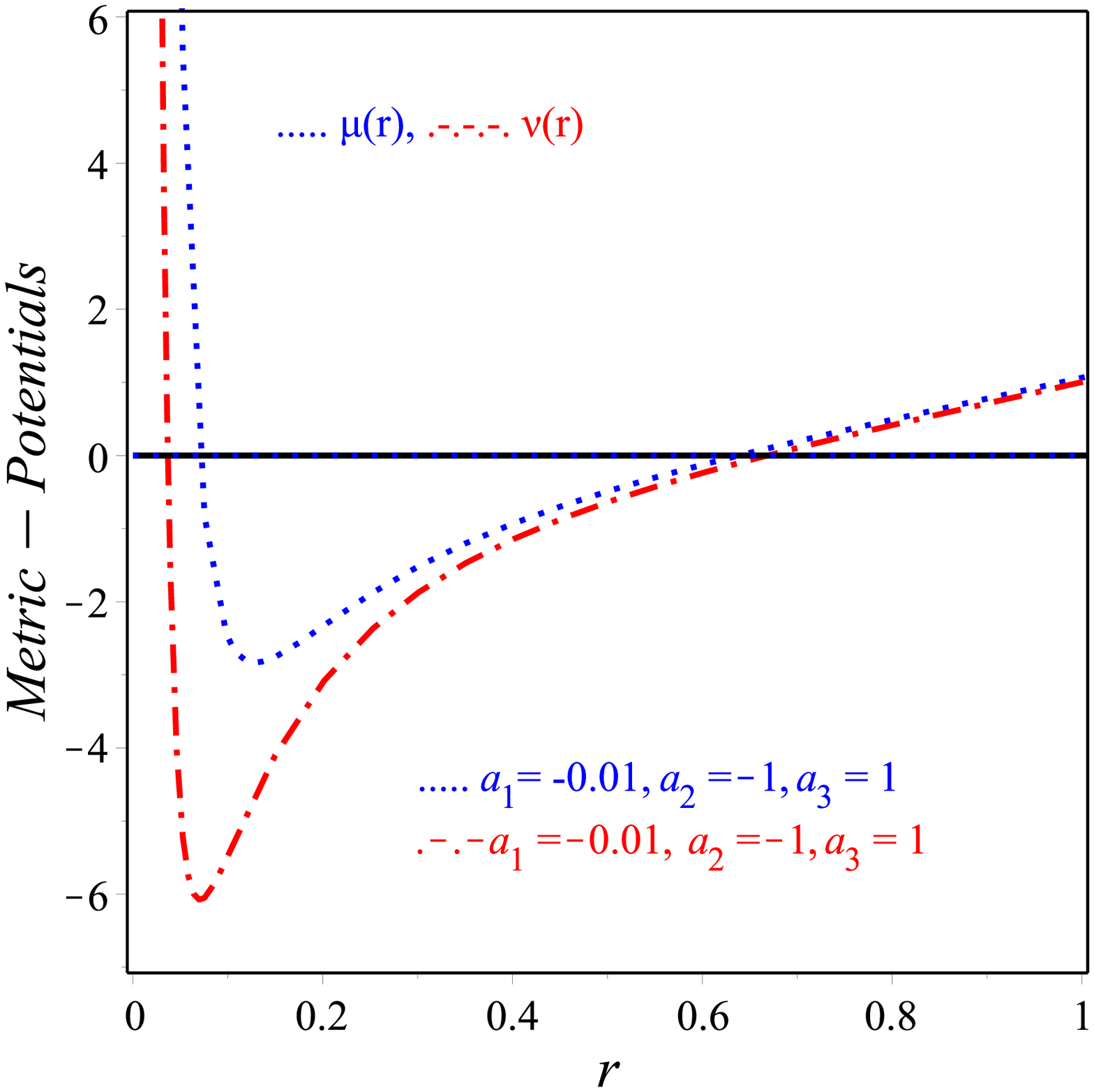}}
\subfigure[~The horizons of the black hole (\ref{mpab1}) of the metric potential $\mu$]{\label{fig:metrdl}\includegraphics[scale=0.23]{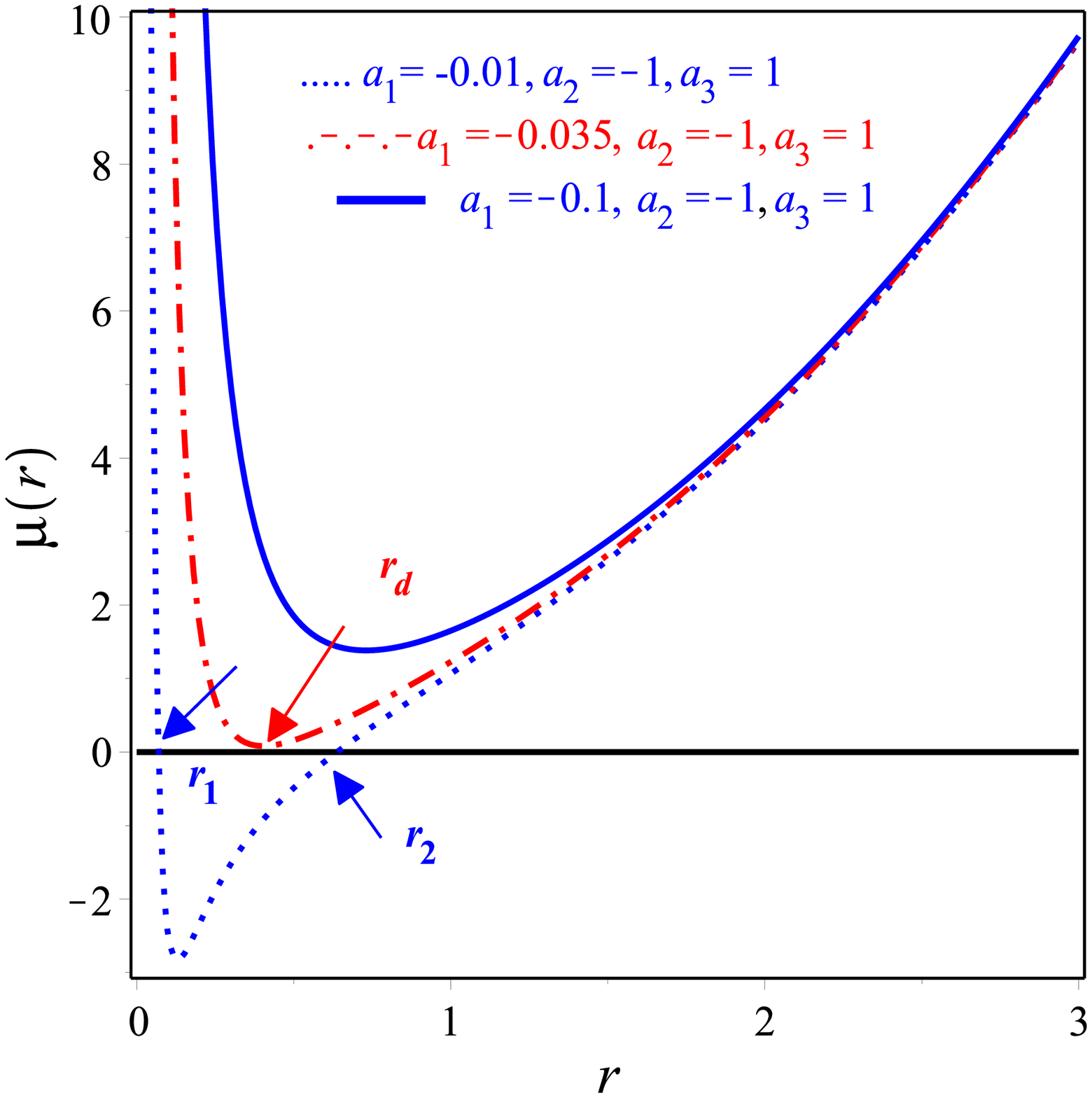}}
\subfigure[~Hawking temperature of black hole (\ref{mpab1})]{\label{fig:templ}\includegraphics[scale=0.23]{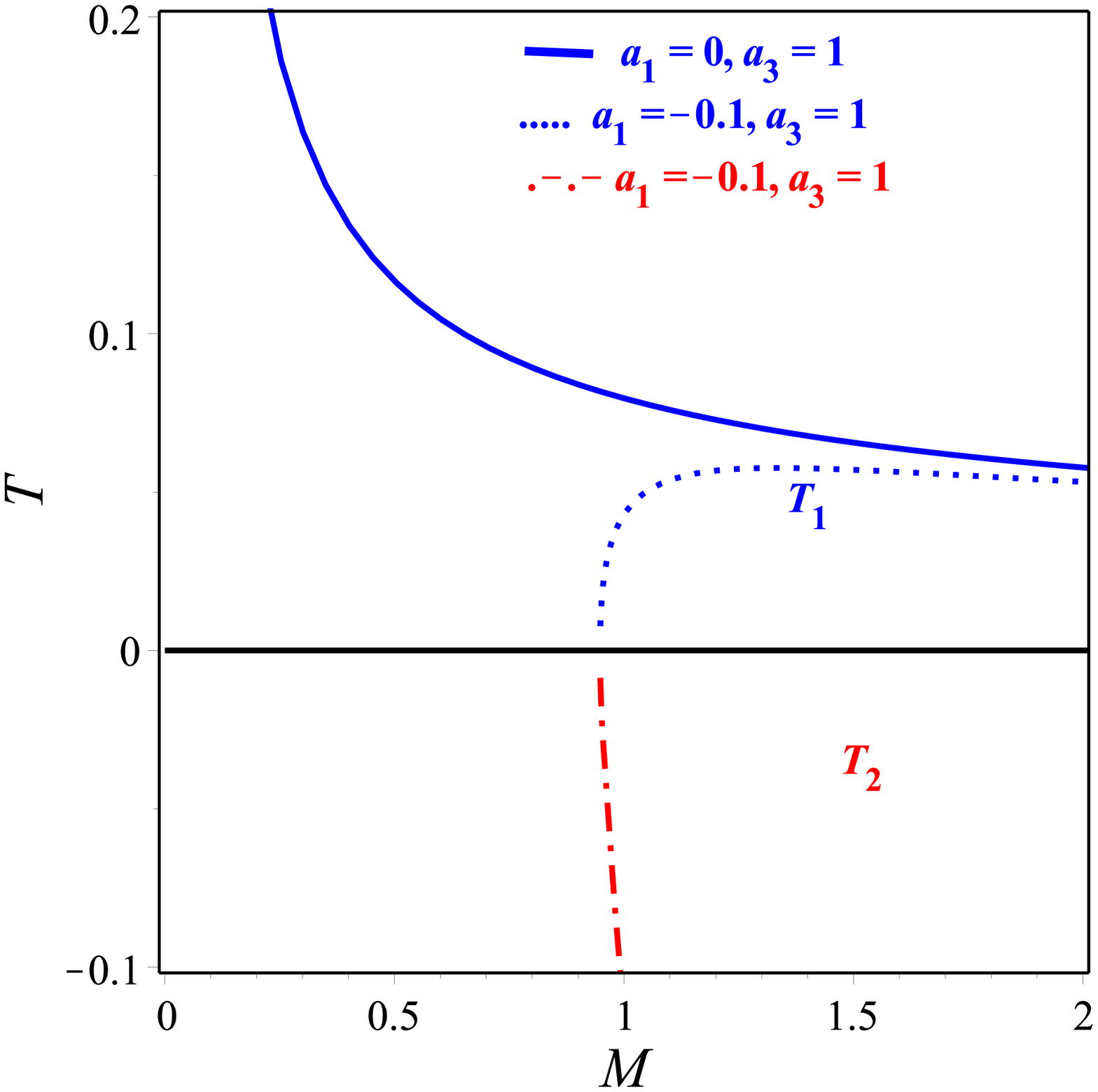}}
\subfigure[~The entropy of the black hole (\ref{mpab1})]{\label{fig:entl}\includegraphics[scale=0.23]{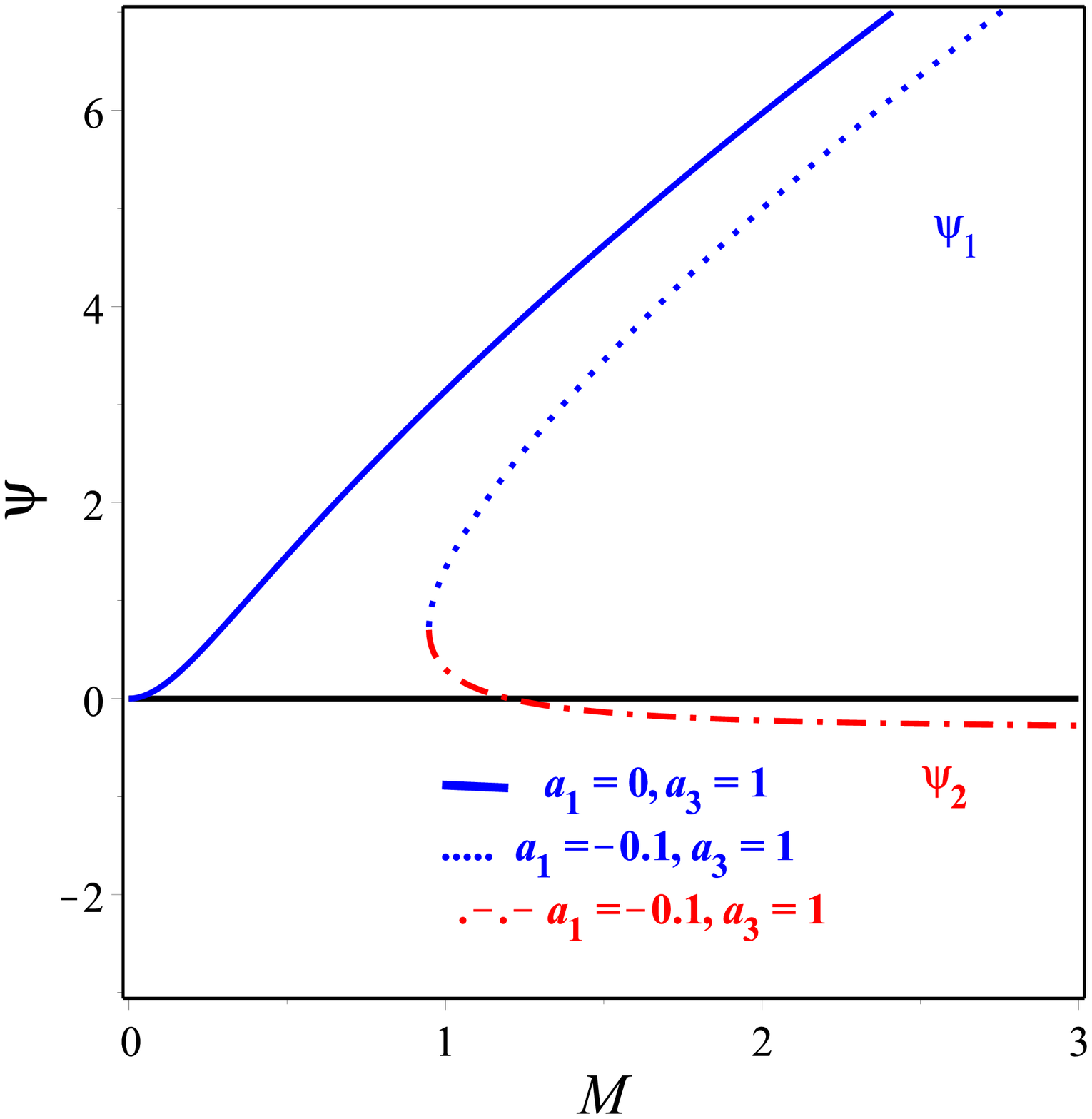}}
\subfigure[~The quasi-local energy of the black hole (\ref{mpab1})]{\label{fig:enrl}\includegraphics[scale=0.23]{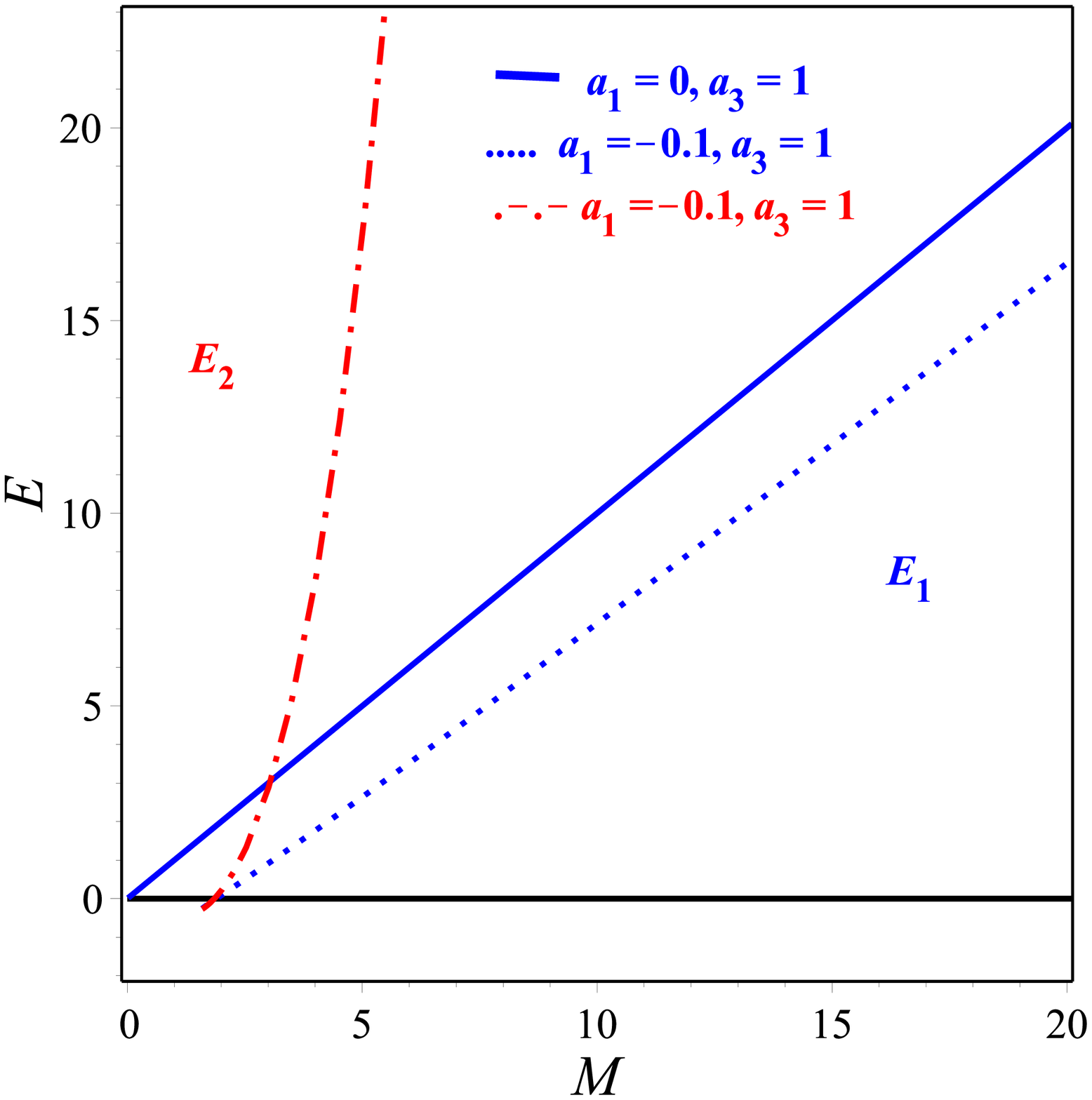}}
\subfigure[~The free energy of the black hole (\ref{mpab1})]{\label{fig:gibl}\includegraphics[scale=0.23]{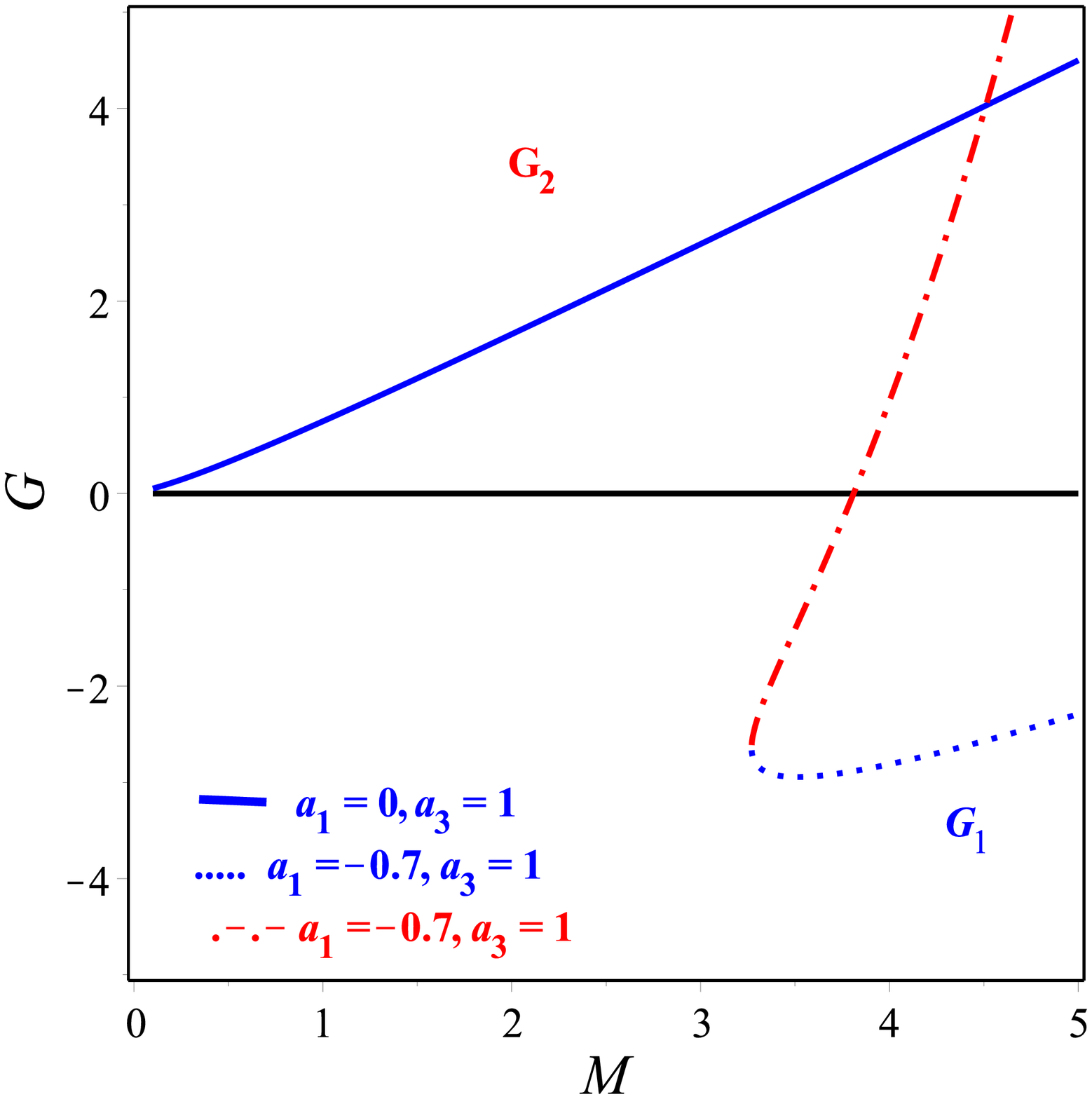}}
\caption[figtopcap]{Schematic plots of thermodynamical quantities of the black hole solution (\ref{mpab1}): \subref{fig:met} shows the behavior
of the metric potentials $\mu$ and $\nu$; \subref{fig:metrd} Typical behaviour of the horizons of the metric potential $\mu(r)$ given
by (\ref{mpab1}); \subref{fig:temp} typical behaviour of the horizon temperature which shows that $T_1$ has a positive decreasing value
while $T_2$ has a negative increasing value; \subref{fig:ent} typical behaviour of the horizon entropy which shows that $ \psi_{1}$ has
an increasing value as $M$ increases while $\psi_{2}$ is positive then becomes negative as $M$ increases; \subref{fig:enr} typical behaviour
of the quasi-local energy that indicates that $E_{(1,2)}$ has positive increasing values and also $E_2>E_1$; \subref{fig:gib} typical behaviour
of the horizon Gibbs free energy which shows that $G_1$ is negative while $G_2$ starts with negative value and as $M$ increase it becomes positive.}
\label{Fig:3}
\end{figure}

%%%%%%%%%%%%%%%%%%%%%%%%%%%%%%%%%%%%%%%%%%%%%%%%%%%%%%%%%%%%%%%%%%%%%%%%%%%%%%%%%%%%%%
In this subsection, we are going to study the black hole (\ref{mpab1}) which is characterized by the mass of the black hole $M$,
the parameter $a_1$ and a positive cosmological effective constant\footnote{The metric potential when $\Lambda_\mathrm{eff.}=a_3$ take the form
\begin{eqnarray}
\label{mpabds1}
&& \mu(r)\approx1+\Lambda_\mathrm{eff} r^2-\frac{2M}{r}-\frac{13a_1}{2r^2}+\cdots\, , \nonumber\\
%\Lambda_\mathrm{eff}r^2+1-\frac{2M}{r}-\frac{a_1{}^2}{r^2}+\frac{8a_1{}^4}{5r^3}+\frac{6a_1{}^2M}{r^3}-\frac{52a_1{}^4}{21r^4}+\cdots\,,\nonumber\\
%
&& \nu(r)\approx 1+\Lambda_\mathrm{eff}(r^2+3a_1)+1-\frac{2M}{r}+\frac{a_1(9a_1{}a_3-7a_1)}{2r^2}+\cdots\,.
% \Lambda_\mathrm{eff}r^2+3-\frac{2M}{r}+\frac{11a_1{}^2}{r^2}+\frac{8a_1{}^4}{5r^3}-\frac{6a_1{}^2M}{r^3}+\frac{452a_1{}^4}{21r^4}+\cdots\,.
%\nonumber
\end{eqnarray}}.
The metric potentials of the black hole (\ref{mpabds1}) are drawn in Fig.~\ref{Fig:3}~\subref{fig:met}.
 From Fig.~\ref{Fig:3}~\subref{fig:met}, we can easily see the two horizons of the metric potentials $\mu(r)$ and $\nu(r)$.
To find the horizons of this black hole, (\ref{mpabds1}), we put $\mu(r)=0$ in Eq.~(\ref{mpabds1}).
This gives four roots two of them are real and the others are imaginary.
These real roots are lengthy however, their behaviors are drawn in Fig.~\ref{Fig:3}~\subref{fig:metrdl}.
It is easy to check that the degenerate horizon for the metric potential $\mu(r)$ given by Eq.~(\ref{mpabds1}) is happened for a specific values
for $\left( a_1,M,r,\Lambda_\mathrm{eff} \right)\equiv(-0.035,0.5,0.3881986644,1)$, respectively which corresponds to the Nariai black hole.
The degenerate behavior is shown is Fig.~\ref{Fig:3}~\subref{fig:metrdl} which shows that the horizon $r_1$ increases with $a_1$ while $r_2$ decreases.
Using Eq.~(\ref{temp}) we draw the behavior of the Hawking temperatures in Fig.~\ref{Fig:3}~\subref{fig:temp} which shows that $T_{1}>T_{2}$.
As Fig.~\ref{Fig:3}~\subref{fig:temp} shows that the $T_1$ has decreasing positive temperature while $T_2$ has increasing negative temperature.

Using Eq.~(\ref{ent}), we draw the entropy in Fig.~\ref{Fig:3}~\subref{fig:entl} which shows increasing value for $\psi_1$ and decreasing value for $\psi_2$.
Using Eq.~(\ref{en}) we calculate the quasi-local energy that is drawn in Fig.~\ref{Fig:3}~\subref{fig:enrl} which also shows positive increasing value for $E_{(1,2)}$.
Finally, we use Eqs.~(\ref{T1}), (\ref{S1}) and (\ref{E1}) in Eq.~(\ref{enr}) to calculate the Gibbs free energies.
The behavior of these free energies are shown in Fig.~\ref{Fig:3}~\subref{fig:gibl} which shows negative increasing value for $G_1$ while $G_2$ starts
with negative value and as $M$ increases it becomes positive.

By the same procedure we can study the case dS, i.e., $\Lambda_\mathrm{eff}a_3<0$ and show the behaviour of their thermodynamical physical quantities.

%%%%%%%%%%%%%%%%%%%%%%%%%%%%%%%%%%%% Section 3 %%%%%%%%%%%%%%%%%%%%%%%%%%%%%%%%%%%%%%%%
\subsection{First law of thermodynamics of the BHs (\ref{mpab}) and (\ref{mpab1})}
%%%%%%%%%%%%%%%%%%%%%%%%%%%%%%%%%%%%%%%%%%%%%%%%%%%%%%%%%%%%%%%%%%%%%%%%%%%%%%%%%%%%%%
It is important for any black hole to check its validity of the first law of thermodynamics. Therefore, we will use the formula of this law in the frame of $f(R)$ as \cite{Zheng_2018}
\begin{equation}
\label{1st}
dE=Td\psi-PdV\, ,
\end{equation}
with $E$, $\psi$, $T$, $P$, and $V$ are the quasi-local energy, the Hawking entropy, the Hawking temperature and the radial component of the stress-energy tensor
that serves as a thermodynamic pressure, $P=T_r{}^r\mid_{\pm}$, and the geometric volume respectively.
The pressure, in the context of $f(R)$ gravitational theory, is figured as \cite{Zheng_2018}
\begin{equation}
P=-\frac{1}{8\pi}\left\{\frac{{F}_1}{r_{\pm}{}^2}+\frac{1}{2}(f({R})-{R}{F}_1)\right\}
+\frac{1}{4}\left(\frac{2{F}_1}
{r_{\pm}}+{F}'_1\right)T\,.
\label{1st}
\end{equation}
For the flat spacetime (\ref{mpab}) if we neglect $O\left(\frac{1}{r^3} \right)$ and $O \left( \frac{1}{r^4} \right)$ to make the calculations more applicable,
we get\footnote{When we neglect the terms of order $O \left( \frac{1}{r^3} \right)$ and $O \left( \frac{1}{r^4} \right)$ and when $A(r)=0$ we get two roots.}
\begin{eqnarray}
\label{1stf}
&&r_{_{(1,2)}}=\frac{2M\pm\sqrt{4M^2+26a_1}}{2}\,,\qquad \psi_{_{(1,2)}}=\frac{\pi \left( 4M^2\pm 2\sqrt{4M^2+26a_1}+15a_1 \right)}{2}\, ,\qquad
T_{_{(1,2)}}=\pm\frac{\sqrt{4M^2+26a_1}}{\pi\, \left( 2M\pm\sqrt{4M^2+26a_1} \right)^2}\, , \nonumber\\
&& E_{_{(1,2)}}=\frac{2 \left( \pm6M\sqrt{4M^2+26a_1} \left[ 8M^2+23a_1 \right] + 48M^4+294a_1M^2+199a_1{}^2 \right)}{3 \left( 2M\pm\sqrt{4M^2+26a_1} \right)^3}\,,\nonumber\\
&& P_{_{(1,2)}}=\frac{4a_1 \left( 10M\sqrt{4M^2+26a_1}+20M^2+69a_1 \right)}{\left( 2M\pm\sqrt{4M^2+26a_1} \right)^4}\,.
\end{eqnarray}
Using (\ref{1stf}) in (\ref{1st}), we can prove that the first law of the flat spacetime is verified.
Repeating the same procedure for the AdS/dS, we get
\begin{eqnarray}
\label{1stf}
&&r_{_{(1,2)}}=\frac{2M\pm\sqrt{4M^2+26a_1}}{2}\,,\qquad \psi_{_{(1,2)}}=\frac{\pi \left( 4M^2\pm 2\sqrt{4M^2+26a_1}+15a_1 \right)}{2}\, ,\qquad
T_{_{(1,2)}}=\pm\frac{\sqrt{4M^2+26a_1}}{\pi\, \left(2M\pm\sqrt{4M^2+26a_1} \right)^2}\, , \nonumber\\
&& E_{_{(1,2)}}=\frac{2 \left( \pm6M\sqrt{4M^2+26a_1} \left[ 8M^2+23a_1 \right] +48M^4+294a_1M^2+199a_1{}^2 \right)}{3 \left( 2M\pm\sqrt{4M^2+26a_1} \right)^3}\,,\nonumber\\
&& P_{_{(1,2)}}= \frac{4a_1 \left( 10M\sqrt{4M^2+26a_1}+20M^2+69a_1 \right)}{\left( 2M\pm\sqrt{4M^2+26a_1} \right)^4}\,.
\end{eqnarray}
%where $\mho=\sqrt{14100m^2b^4-75b^6-3600m^4b^2+8640mb^6+1296b^8-960m^3b^4}$.
Substituting the form of $r$ into the thermodynamical quantities we get
\begin{eqnarray}
\label{1stv}
&& dE=\frac{4m^4+b^2m^2+b^4}{8\,m^4}\,,\qquad
\psi=\frac{\pi \left( 40m^4-20m^2b^2-8mb^4-25b^4 \right)}{10\,m^3}\, , \qquad
T=\frac{4mb^4-20m^4+35b^4+15m^2b^2}{160\,\pi\, m^5}\,,\nonumber\\
&& P=-\frac{b^2 \left( 30m^2+75b^2+8mb^2 \right)}{640\, \pi \,m^6}\,.
\end{eqnarray}
If we use Eq.~(\ref{1stv}) in (\ref{1st}), we can verify the first law of thermodynamics for the BH (\ref{mpab}).

If one repeats the same procedure for the BH (\ref{mpab1}), one can verify the first law of thermodynamics provided that we neglect all the quantities
containing $b$ to make the calculation easier to carry out.

\section{Discussion and conclusions }\label{S77}
It is well-known that spherically symmetric spacetime is an essential spacetime in GR for many reasons:
First because it is the easiest geometry that one can start to use in the study of GR.
In this exposition, we tackle the problem of applying the charged field equations of $f(R)$ to a spherically symmetric spacetime with two unknown functions.
We derive a highly nonlinear five differential equations in four unknowns, two of the metric potential, the third is the derivative of $f(R)$,
and the fourth is the gauge potential of the Maxwell field.
We study some limits of this system of differential equations to test its compatibility with the previous studies and got consistent results of BHs
derived for constant and non-constant Ricci scalar.
Then we turn our attention to the general case and solve this system by assuming a specific form of the derivative of $f(R)$.
This assumption involves one constant that when it equals zero we got the GR theory limit.
We derive the form of the metric potential and the gauge potential of the Maxwell field and show they depend on the convolution and error functions respectively.
Under some constraints, if these functions are set zero, we got to the Schwarzschild BH of GR which means that in the limit of higher-order curvature theory,
we cannot generate the Reissner-Nordstr\"om BH.
This gives us an indication that the effect of the higher-order curvature terms restored in the convolution and error functions.

To gain some physics of these BHs we studied their asymptotes and show that we have a constant of integration that plays an important role
in the asymptote, i.e., when this constant is vanishing/non-vanishing, we got two different asymptotes of the BHs, a flat and AdS/dS spacetimes respectively.
Moreover, we studied the invariants of these BHs and showed that their singularities are milder than those of GR BH.
By the use of the geodesic deviation, we derived the conditions of stability of those BHs and showed their stable areas as shown in Figs.~\ref{Fig:1}~\subref{fig:1a} and \subref{fig:1b}.
Also, we have calculated some thermodynamics quantities like the Hawking temperature, entropy, quasi-local energy, and the Gibbs free energy to know the behavior
of those BHs analytic and graphically and show in detail that such BHs satisfy the first law of thermodynamics.
Finally, if we use the results of odd perturbation presented in \cite{Elizalde:2020icc}, we can show easily that the BHs derived in this study are stable.

To conclude, we have derived new charged BHs in the context of $f(R)$ that their Ricci scalars are not constant.
The originality of those BHs comes from the constant that involves the assumption of the derivative function of $f(R)$.
If we follow the same procedure but with another assumption of the derivative function of $f(R)$, we can get new BHs but this needs more study because of the complication of the system.
This will be done elsewhere.

\newpage
%%%%%%%%%%%%%%%%%%%%%%%%%%%% Section 7 %%%%%%%%%%%%%%%%%%%%%%%%%%%%%
 {
 \begin{center}
{\bf{Appendix A}}
\end{center}
%\begin{center}
{\centerline{\bf The form of Kretschmann scalar }}
%\end{center}}\vspace{0.3cm}
 \renewcommand{\eqref}{}
%%%%%%%%%%%%%%%%%%%%%%%%%%%%%%%%%%%%%%%%%%%%%%%%%%%%%%%%%%%%%%%%%%%%
The exact form of the Kretschmann
scalar is given as:
\begin{eqnarray*}
&& R_{\mu \nu \rho \sigma} R^{\mu \nu \rho \sigma}=\frac{1}{\left( {r}^{2}+a_1 \right)^{2}{r}^{10}}\Bigg\{ 16 \e^{{\frac {3a_1}{{r}^{2}}}}\Bigg[3\left( 3{a_1}^{4}+12\,{a_1}^{2}{r}^{4}+6\,{a_1}^{3}{r}^{2}-2a_1{r}^{6}+{r}^{8} \right){H}^{2}-4{H}_2 \left( {r}^{2}a_1 +{r}^{4}+6{a_1}^{2} \right) {r}^{2}a_1H+4\times \nonumber\\
 && H_2{}^{2}{a_1}^{2}\left(2{a_1}^{2}+2{r}^{2}a_1+{r}^{4} \right) \Bigg] \left( \int \! \frac{{H}_1 {\e^{-{\frac {3a_1}{2{r}^{2}}}}} \left( {r}^{2}+13a_1 \right) {r}^{2}}{\left( {r}^{2}+a_1 \right) \left( 2a_1[{H}_1{H}_2-{H}{H}_3] -3{H}_1 {H} {r}^{2} \right)}{dr} \right)^{2}-16 \Bigg[\Bigg\{ \Bigg(18 \left( 2{a_1}^{2}{r}^{2}+2{r}^{4}a_1-{r}^{6}+{a_1}^{3} \right) {H}_1\nonumber\\
 && -2\left({r}^{2}a_1+{r}^{4}+18{a_1}^{2} \right) {r}^ {2}H_3 \Bigg) {H} +2{H}_2 \left( \left(2{r}^{6}+5{r}^{4}a_1 \right) {H}_1 +2a_1{H}_3 \left(2 {a_1}^{2}+2{r}^{2}a_1+{r}^{4} \right) \right) \Bigg\} {r}^{3}a_1{\e^{{\frac 3{a_1}{2{r}^{2}}}}}\nonumber\\
 &&\times\int \!\frac{{\e^ {-{\frac {3a_1}{2{r}^{2}}}}}{H}\left( {r}^{2}+13\,a_1 \right)}{ {r}\left( {r}^{2}+a_1 \right) \left( 2a_1[{H}_1 {H}_2-{H}{H}_3] -3\,{H}_1 {H} {r}^{2} \right)}{dr}+ 3\left( 3{a_1}^{4}+12{a_1}^{2}{r}^{4}+6{a_1}^{3}{r}^{2} -2a_1{r}^{6}+{r}^{8} \right) a_2{\e^{{\frac {3a_1}{2{r}^{2}}}}} {H}{}^{2}\nonumber\\
 &&+ \Bigg\{9 a_1ra_3\,{ \e^{{\frac {3a_1}{2{r}^{2}}}}} \left( 2\,{a_1}^{2}{r}^{2}+2\,{r}^{ 4}a_1-{r}^{6}+{a_1}^{3} \right) {H}_1 -2\left( {r}^{2}a_1+ {r}^{4}+18{a_1}^{2} \right) {r}^{3}a_1{\e^{{\frac {3a_1}{2{r}^{2}}}}}a_3\,{H}_3 -4 \left({r}^{2}a_1+{r}^{4}+18{ a_1}^{2} \right) a_2\,a_1{\e^{{\frac {3a_1}{2{r}^{2}}}}}{H}_2\nonumber\\
 &&
 +{r}^{3} \left( {\e^{-3/2\,{\frac {a_1}{{ r}^{2}}}}} \left(5 a_1-{r}^{2} \right) \left({r}^{2}+13a_1 \right) {\e^{{\frac {3a_1}{2{r}^{2}}}}}-\left( {r}^{2}+a_1 \right) ^{2} \right) \Bigg\}{r}^{2}{H}+2{H}_2 \Bigg[ \left(5a_1+2{r}^{2} \right) {r}^{7}a_3\,{H}_1 + \Bigg\{2{r}^{3}a_3\left( 2{a_1}^{2}+2{r}^{2}a_1+{r}^{4} \right) {H}_3 \nonumber\\
 &&+a_2\, \left( 2{a_1}^{2}+2{r}^{2}a_1+{r}^{4} \right) {H}_2 -{\e^{-3/2\,{\frac {a_1}{{r}^{2}}}}}{r}^{3} \left( {r}^{2}+13a_1 \right) \Bigg\} a_1 \Bigg] a_1{\e^{{\frac {3a_1}{2{ r}^{2}}}}} \Bigg] {\e^{{\frac 3{a_1}{2{r}^{2}}}}}\int \!\frac{{H}_1 {\e^{-{\frac {3a_1}{2{r}^{2}}}}} \left( {r}^{2}+13\,a_1 \right) {r}^{2}}{ \left( {r}^{2}+a_1 \right)\left( 2\,{H}_1 a_1{H}_2 -3\,{H}{H}_1 {r}^{2}-2\,{H} a_1{H}_3 \right)}{dr} \nonumber\\
 &&+16 \left( 3 \left( 2{r}^{8}+6\,{a_1}^{ 3}{r}^{2}+3{a_1}^{4}+6\,{a_1}^{2}{r}^{4}+2/3\,a_1{r}^{6} \right)H_1{}^{2}+4 \left(5a_1+2{r}^{2} \right) {r}^{4}a_1H_3 H_1 +4{a_1}^{2}{H}_3{}^{2} \left(2{a_1}^{2}+2{r}^{2}a_1+{r}^{ 4} \right) \right) {r}^{6}{\e^{{\frac {3a_1}{{r}^{2}}} }} \nonumber\\
 &&\times \left( \int \!\frac{\frac{\e^{{\frac {3a_1}{2{r}^{2}}}}}H \left( {r}^{2}+13\,a_1 \right)}{ r \left( {r}^{2}+a_1 \right) \left( 2a_1[H_1 H_2-H H_3] -3\,{H}_1H{r}^{2} \right)}{dr} \right) ^{2}+16{r}^{3}{\e^{{\frac {3a_1}{2{r}^{2}}}}} \Bigg\{ a_2\, \Bigg[ 9 \left( 2\,{a_1}^{2}{r}^{2}+2\,{r}^{4}a_1-{r}^ {6}+{a_1}^{3} \right) H_1 -2 {r}^{2}H_3\Bigg\{{r}^{2}a_1 \nonumber\\
 &&+{r}^{4} +18 {a_1}^{2} \Bigg\} \Bigg] a_1{\e^{{\frac {3 a_1}{2{r}^{2}}}}}H +3{r}^{3} \left( 2{r}^{8}+2{a_1}^{3}{r}^{2}+{ a_1}^{4}+6{a_1}^{2}{r}^{4}+2a_1{r}^{6} \right) {\e^{{\frac { 3a_1}{2{r}^{2}}}}}a_3H_1{}^{2}+{r}^{4} \Bigg\{ 2 \left(5 a_1+2{r}^{2} \right) a_1{\e^{{\frac {3a_1}{{2r}^{2}}}}}[2a_3H_3r^3+a_2H_2] \nonumber\\
 &&\left( {\e^{-{\frac {3a_1}{2{r}^ {2}}}}} \left(2 a_1-{r}^{2} \right) \left({r}^{2}+13a_1 \right) {\e^{3/2\,{\frac {a_1}{{r}^{2}}}}}- \left( {r}^{2}+a_1 \right) ^{2} \right) r \Bigg\}H_1 +\Bigg\{ 4{r}^{3}a_3\, \left(2 {a_1}^{2}+2{r}^{2}a_1+{r}^{4} \right) {H}_3 +4a_2\, \left( 2{a_1}^{2}+2{r}^{2}a_1+{r}^{4} \right) {H}_2\nonumber\\
 &&
 -{\e^{-{\frac {3a_1}{2{r}^{2}}}}}{r}^{ 3} \left({r}^{2}+13a_1 \right) \Bigg\}{a_1}^{2}{\e^{{ \frac {3a_1}{2{r}^{2}}}}}H_3 \Bigg\} \int \!\frac{{\e^{-{ \frac {3a_1}{2{r}^{2}}}}}H \left( {r}^{2}+13\,a_1 \right)}{ {r} \left( {r}^{2}+a_1 \right)\left( 2\,H_1 a_1{H}_2 -3 \,HH_1 {r}^{2}-2\,Ha_1H_3 \right)}{dr }\nonumber\\
 && +12 \left(3{a_1}^{4}+6{a_1}^{2}{r}^{4}+6\,{a_1}^{3}{r}^{2}-2a_1{r}^{6 }+{r}^{8} \right) {a_2}^{2}{\e^{{\frac {3 a_1}{{r}^{2}}}}} H^{2}+8a_2\,{r}^{2} \Bigg\{9 a_1ra_3\,{\e^{{\frac {3a_1}{2{r}^{2}} }}} \left( 2\,{a_1}^{2}{r}^{2}+2\,{r}^{4}a_1-{r}^{6}+{a_1}^{3} \right)H_1 -2\Bigg[{r}^{2}a_1+{r}^{4}\nonumber\\
 &&
 +18{a_1}^{2} \Bigg] {r}^ {3}a_1{\e^{{\frac {3a_1}{2{r}^{2}}}}}a_3\,H_3 -2\left( {r}^{2}a_1+{r}^{4}+18{a_1}^{2} \right) a_2\,a_1{ \e^{{\frac {3a_1}{2{r}^{2}}}}}H_2 +\,{r} ^{3} \Bigg[ {\e^{-{\frac {3a_1}{2{r}^{2}}}}} \left(5a_1-{r}^{ 2} \right) \left( {r}^{2}+13a_1 \right) {\e^{{\frac {3a_1}{2{ r}^{2}}}}} - \left( {r}^{2}+a_1 \right) ^{2} \Bigg] \Bigg\} {\e^{{\frac {3a_1}{2{r}^{2}}}}}H\nonumber\\
 && +12{r}^{6} \left( 2{r}^{8}+6\,{a_1}^{3}{r}^{2}+3{a_1}^{4}+6\,{a_1}^{2}{r}^{4}+ a_1{r}^{6} \right){\e^{{\frac {3a_1}{{r}^{2}}}}}{a_3}^{2} H_1{}^{2}+8{r}^{7 }{\e^{{\frac {3a_1}{2{r}^{2}}}}} \Bigg[ 2\left(5 a_1+{r}^{2} \right) {r}^{3}a_1{\e^{{\frac {3a_1}{2{r}^{2}}}}}[r^3a_3\,H_3 + a_2H_2]\nonumber\\
 && + \left( { \e^{-{\frac {3a_1}{2{r}^{2}}}}} \left(2 a_1-{r}^{2} \right) \left({r}^{2}+13a_1 \right) {\e^{{\frac {3a_1}{2{r}^{2}}}}}- \left( {r}^{2}+a_1 \right)^{2} \right) r \Bigg] a_3\,H_1 +16{r}^{6}{a_1}^{2} {\e^{{\frac {3a_1}{{r}^{2}} }}}{a_3}^{2} \left(2 {a_1}^{2}+2{r}^{2}a_1+{r}^{4} \right) \left( H_3 \right) ^{2}\nonumber\\
 && +16\left(2 a_2\, \left( 2{a_1}^{2}+2{r}^{2}a_1+{r}^{4} \right) H_2 -{\e^{-{\frac {13a_1}{2{r}^{2}}}}}{r}^{3} \left( {r}^ {2}+13a_1 \right) \right) {r}^{3}{a_1}^{2} {\e^{{\frac {3a_1} {{r}^{2}}}}}a_3\,H_3 +16\,{a_2}^{2 }{a_1}^{2}{\e^{{\frac {3a_1}{2{r}^{2}}}}} \left( 2{a_1}^{2}+2{r}^{2}a_1+{r}^{4} \right)H_2{}^{2}\nonumber\\
 &&
 -16{\e^{-3/2\,{\frac {a_1}{{r}^{2}}}}}a_2\,{ r}^{3}{a_1}^{2}{\e^{{\frac {3a_1}{{r}^{2}}}}} \left({r}^{2}+13a_1 \right) H_2+4 \left( { \e^{-{\frac {3a_1}{{r}^{2}}}}} \left( {r}^{2 }+13a_1 \right) ^{2}{\e^{{3\frac {a_1}{{r}^{2}}}}}+ \left( {r}^{2}+a_1 \right) ^{2} \right) {r}^{6} \Bigg\}\,. \nonumber\\
 \end{eqnarray*}
The above equation shows that the Kretschmann
scalar has a true singularity at the origin.
By the same method one can show that the squared of the Ricci tensor
and the Ricci scalar have a singularity at the origin.}
%%%%%%%%%%%%%%%%%%%%%%%%%%%%%%%%%
%\bibliographystyle{apsrev}
%\bibliography{JRPHSRef}

\end{document}